\documentclass[fleqn,usenatbib]{mnras}

\usepackage{newtxtext,newtxmath}

\usepackage[T1]{fontenc}
\usepackage{ae,aecompl}
\usepackage{CJKutf8}
\usepackage{romannum}

\hypersetup{draft}

\usepackage{graphicx}	
\usepackage{amsmath}	
\usepackage{amssymb}	

\usepackage{lipsum}

\usepackage{tikz}
\usetikzlibrary{positioning}

\newcommand{\HII}{\ion{H}{ii}}
\newcommand{\te}{$T_e$}

\title[Calibrate strong-line abundance using ANN]{A Machine Learning Artificial Neural Network Calibration of the Strong-Line Oxygen Abundance}
\author[Ho]{
I-Ting Ho (何宜庭),$^{1}$\thanks{E-mail: iting@mpia.de}
\\
$^{1}$Max Planck Institute for Astronomy, K\"{o}nigstuhl 17, 69117 Heidelberg, Germany\\
}

\date{Accepted XXX. Received YYY; in original form ZZZ}

\pubyear{????}

\begin{document}
\begin{CJK*}{UTF8}{bkai}
\label{firstpage}
\pagerange{\pageref{firstpage}--\pageref{lastpage}}
\maketitle

\begin{abstract}

The \HII\ region oxygen abundance is a key observable for studying chemical properties of galaxies. Deriving oxygen abundances using optical spectra often relies on empirical strong-line calibrations calibrated to the direct method. Existing calibrations usually adopt linear or polynomial functions to describe the non-linear relationships between strong line ratios and \te\ oxygen abundances. Here, I explore the possibility of using an artificial neural network model to construct a non-linear strong-line calibration. Using about 950 literature \HII\ region spectra with auroral line detections, I build multi-layer perceptron models under the machine learning framework of training and testing. I show that complex models, like the neural network, are preferred at the current sample size and can better predict oxygen abundance than simple linear models. I demonstrate that the new calibration can reproduce metallicity gradients in nearby galaxies and the mass-metallicity relationship. Finally, I discuss the prospects of developing new neural network calibrations using forthcoming large samples of \HII\ region and also the challenges faced. 
\end{abstract}




\section{Introduction}

The \HII\ region oxygen abundance, a proxy for the metallicity of the interstellar medium (ISM), is a key quantity for studying chemical evolution of galaxies. Oxygen is the most common heavy element in the ISM (after helium and lithium). Together with other relatively abundant elements (e.g. nitrogen, sulfur), they produce a wealth of emission lines in the optical wavelength which are useful for constraining the ISM chemical abundances. Observing these optical emission lines has allowed us to significantly advance our understanding of fundamental chemical characteristics of star-forming galaxies, such as the oxygen abundance gradient \citep[e.g.][]{Searle:1971kx,Zaritsky:1994lr,Sanchez:2014fk,Ho:2015hl} and the mass-metallicity relationship \citep[e.g.][]{Lequeux:1979lr,Tremonti:2004fk,Zahid:2013kx}.

Although observing emission lines from \HII\ regions is relatively straight-forward, deriving accurate oxygen abundances is non-trivial. A variety of methods is available in the literature (see reviews by \citealt{Stasinska:2004aa} and \citealt{Peimbert:2017aa}); however, large systematic differences between different methods exist and the disagreement remains unresolved \citep{Kewley:2008qy}. One of the more accurate ways is through measuring the electron temperature (\te) and density ($n_e$) of the line-emitting gas (see \citealt{Perez-Montero:2017aa} for a tutorial). This method, known as the direct method, is built on the physical basis that the emissivity of collisionally excited lines (CELs) depends strongly on \te. When \te\ and $n_e$ are known, ionic abundances can be inferred from CEL to hydrogen recombination line ratios. The total elemental abundance can subsequently be derived by accounting for unseen stages of ionisation. In this method, $n_e$ can be easily determined using the density-sensitive doublets [\ion{O}{ii}]$\lambda\lambda3726,3729$ and [\ion{S}{ii}]$\lambda\lambda6717,6731$. However, constraining \te\ requires detecting auroral lines originating from transitions between upper energy levels, e.g.~[\ion{O}{iii}]$\lambda4363$, [\ion{N}{ii}]$\lambda5755$. A key limitation of the direct method comes from the faintness of the auroral lines, typically 2 orders of magnitude dimmer than the H$\beta$ line. 

It is worth noting that the direct method is not perfect and has an important caveat. Oxygen abundances estimated base on recombination lines (RLs) are almost always higher than those from the CELs, by about 0.2 to 0.3~dex \citep[e.g.][]{Peimbert:2003aa,Esteban:2002aa,Esteban:2009aa,Garcia-Rojas:2007aa}. One possible origin of this so-called ``abundance discrepancy factor'' problem is the presence of significant temperature, density, and possibly chemical composition fluctuation inside nebulae , which affect the direct method morer than the RL method . The CELs are expected to originate from relatively low metallicity, high \te, and low density regions inside a nebula, leading to an underestimate of the overall metallicity (see \citealt{Peimbert:2017aa} for a review).  Recent spectroscopic studies of blue and red supergiant stars in nearby galaxies, however, report stellar oxygen abundances more consistent with gas-phase abundances measured from the direct method than the RLs (\citealt{Bresolin:2016aa} and references therein).


The faintness of the auroral lines severely limits its practical use when observations do not achieve excellent signal-to-noise ratios. Empirical strong-line calibrations are developed and have gained popularity among those using shallower (spectroscopic survey) data to study chemical properties of galaxies. Strong line indexes sensitive to metallicity, such as $\rm R_{23}$\footnote{$\rm R_{23}\equiv([\ion{O}{ii}] + [\ion{O}{iii}]\lambda\lambda4959,5007)/H\beta$} \citep{Pilyugin:2001aa,Pilyugin:2005aa}, O3N2\footnote{$\rm O3N2\equiv\log(\frac{[\ion{O}{iii}]/H\beta}{[\ion{N}{ii}]/H\alpha})$}, N2\footnote{$\rm N2\equiv\log([\ion{N}{ii}]/H\alpha)$}  \citep{Pettini:2004lr,Marino:2013vn}, $\rm O_{32}$\footnote{$\rm O_{32}\equiv[\ion{O}{iii}]/[\ion{O}{ii}]$} \citep{Maiolino:2008aa,Curti:2017aa}, have been calibrated to the \te\ metallicity. In general,  \te\ metallicities derived from high quality \HII\ spectra (or in some cases integrated galaxy spectra) are plotted against a selected strong line index. A simple linear or polynomial function is fit to the data and then reported as an empirical calibration. 

\begin{figure}
\includegraphics[width = \columnwidth]{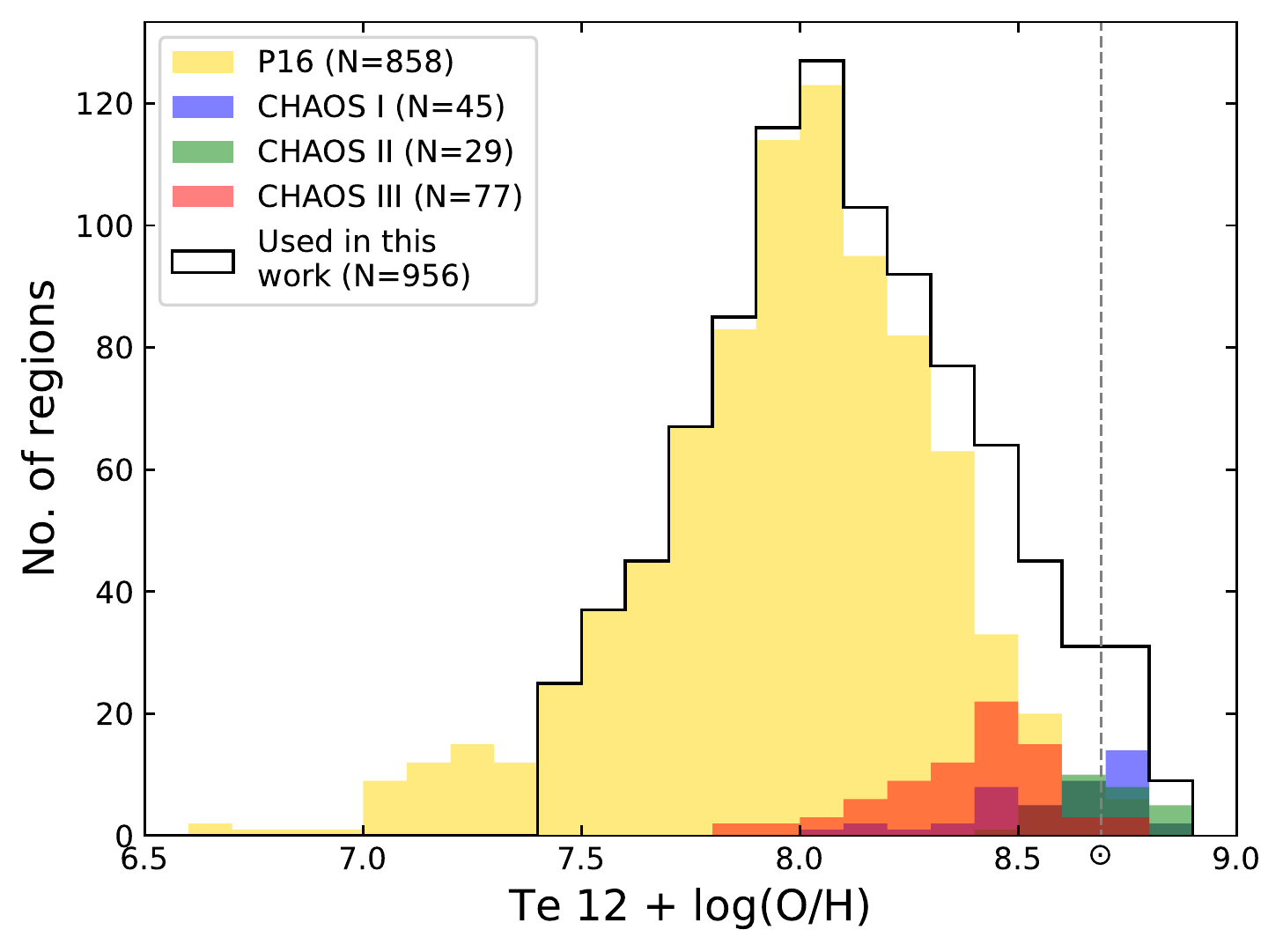}
\caption{\te\ oxygen abundance distributions of different samples. Literature \HII\ region data compiled by \citet[P16]{Pilyugin:2016aa} are shown in yellow. Data from the CHAOS Project are shown in blue \citep[][CHAOS~\Romannum{1}]{Berg:2015aa}, green \citep[][CHAOS~\Romannum{2}]{Croxall:2015aa}, and red \citep[][CHAOS~\Romannum{3}]{Croxall:2016tg}. The black solid line indicates the working sample adopted in this work. }\label{fig_sample}
\end{figure}

The simplest cases are the linear calibrations of the O3N2 and N2 indexes. For the $\rm R_{23}$ diagnostic, more complex functions are needed because the $\rm R_{23}$ index is known to be double valued in metallicity \citep[e.g.][]{Pagel:1979aa,Pagel:1980aa,Kobulnicky:2004lr}. In this situation, either two functions are fit separately to the upper and lower branches or a high-order polynomial is adopted. Photoionisation models, however, suggest that all these indexes have secondary dependency with the ionisation parameter \citep[e.g.][]{Kewley:2002fj,Dopita:2013qy}. It is likely that, by including more line ratios, the calibration dispersion can be reduced.

It is non-trivial to include additional line ratios into the calibrations and find suitable analytical formulae. 
Steps moving from 1- or 2-dimensional calibrations (where one or two line ratios are used) to high-dimensional calibrations (where multiple line ratios are fit simultaneously) have been taken in various work \citep[e.g.][]{Maiolino:2008aa,Curti:2017aa}. A particularly efficient approach was taken by \citet{Pilyugin:2016aa}. 
\citeauthor{Pilyugin:2016aa} recognised that the metallicity-dependent index N2 exhibits a secondary correlation with the excitation parameter P\footnote{$\rm P\equiv[\ion{O}{iii}]\lambda\lambda4959,5007/([\ion{O}{ii}] + [\ion{O}{iii}]\lambda\lambda4959,5007)$}, particularly at high \te\ (low metallicities).  By carefully designing the fitting formulae, \citet{Pilyugin:2016aa} were able to develope two 3-dimensional calibrations: an R calibration using [\ion{O}{ii}], [\ion{O}{iii}] and [\ion{N}{ii}], and an S calibration using [\ion{O}{iii}], [\ion{N}{ii}] and [\ion{S}{ii}].

Another shortcoming of fitting one index at a time is that different indexes are sensitive to metallicity in different ranges, and the scatters could also be different. When multiple diagnostic indexes are available in a spectrum, different metallicities can be reported by using different indexes (and calibrations). A practical approach is to combine the different metallicities by taking the mean or  minimising the $\chi^2$ in the space defined by the selected diagnostics \citep{Curti:2017aa,Maiolino:2008aa}. 

Calibrating strong-line to \te\ metallicities is fundamentally searching for a function that can map the measured line ratios to \te\ metallicities using a set of well-characterised sample. This function is likely to have a high dimensionality \citep[e.g.][]{Vogt:2014aa}. The linear or polynomial fits in the literature describe only projections of this complex function onto certain index space. 
Here, I present an exploratory work of using a neural network model to recover the underlying function. The main goal of this work is to examine the possibility of fitting an arbitrarily complex, non-linear model to existing data (approximately 950 \HII\ regions) and improving the calibration accuracy by directly ``learning'' from the data. By performing supervised learning under a machine learning framework, I will ``train'' a strong line calibration that is data-driven and has no built-in physics. I will show that my first-try model can deliver reasonably good performance, even exceeding most popular literature strong-line calibrations. I will demonstrate that the neural network model can capture line ratios that are known to be sensitive to metallicity in photoionisation models, and successfully reproduce the well-characterised metallicity gradient and mass-metallicity relationship. I will end the article by discussing the prospect of this new approach and some challenges faced ahead. 
\begin{figure}
\includegraphics[width = \columnwidth]{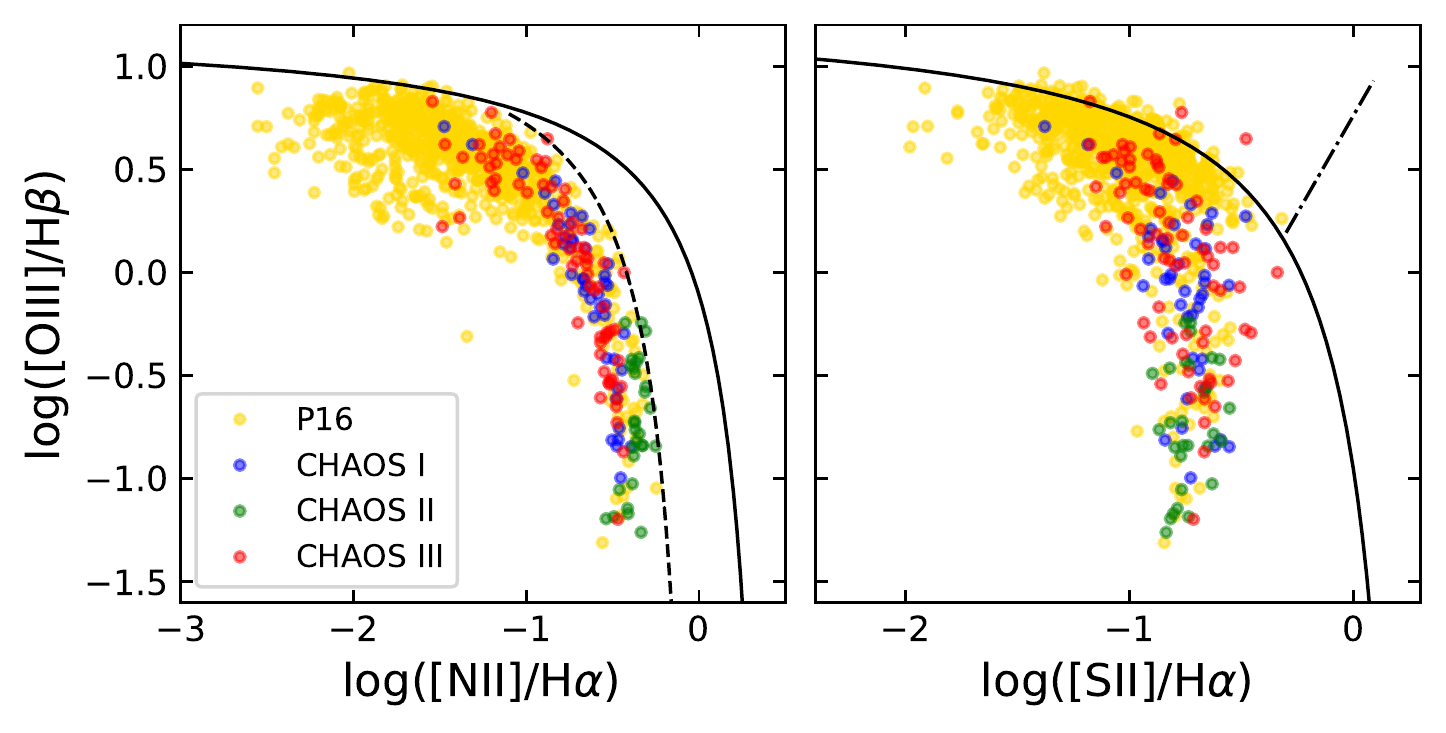}
\caption{Line-ratio diagrams of the working sample (black histogram in Figure~\ref{fig_sample}. The different colors correspond to the different sources as indicated in the legend. }\label{fig_bpt}
\end{figure}

Through out the paper, the following convention is followed: [\ion{O}{ii}] = [\ion{O}{ii}]$\lambda\lambda3726,3729$, [\ion{O}{iii}] = [\ion{O}{iii}]$\lambda5007$, [\ion{N}{ii}] = [\ion{N}{ii}]$\lambda6583$, and [\ion{S}{ii}] = [\ion{S}{ii}]$\lambda\lambda6716,6731$. The best-fit neural network model and corresponding  \textsc{python} wrapper script is publicly available at \url{https://github.com/hoiting/OxygenMLP}.The fluxes and oxygen abundances (as well as their original references) adopted in training and testing are also stored in the online repository.

\begin{figure*}
\tikzset{%
  every neuron/.style={
    circle,
    draw,
    minimum size=0.8cm
  },
  neuron missing/.style={
    draw=none, 
    scale=2.5,
    text height=0.333cm,
    execute at begin node=\color{black}$\vdots$
  }
}

\begin{tikzpicture}[x=1.8cm, y=1.cm, >=stealth]

\foreach \m/\l [count=\y] in {1,2,3,missing,4}
  \node [every neuron/.try, neuron \m/.try] (input-\m) at (0,2.5-\y) {};

\foreach \m [count=\y] in {1,missing,2}
  \node [every neuron/.try, neuron \m/.try ] (hidden1-\m) at (1.33,2-\y*1.25) {};

\foreach \m [count=\y] in {1,missing,2}
  \node [every neuron/.try, neuron \m/.try ] (hidden2-\m) at (2.66,2-\y*1.25) {};

\foreach \m [count=\y] in {1}
  \node [every neuron/.try, neuron \m/.try ] (output-\m) at (4,.5-\y) {};

\foreach \l [count=\i] in {1,2,3,16}
  \draw [<-] (input-\i) -- ++(-1,0)
    node [above, midway] {Feature {\l}};


  \node [] (input) at (0,2.5-1) {S};
  \node [] (input) at (0,2.5-2) {S};
  \node [] (input) at (0,2.5-3) {S};
  \node [] (input) at (0,2.5-5) {S};

  \node [] () at (1.33,2-1*1.25) {$\Sigma$|R};
  \node [] () at (1.33,2-3*1.25) {$\Sigma$|R};

  \node [] () at (2.66,2-1*1.25) {$\Sigma$|R};
  \node [] () at (2.66,2-3*1.25) {$\Sigma$|R};

  \node [] () at (4,0.5-1) {$\Sigma$};
\node [every neuron/.try, neuron1/.try] () at (6,+2) {S};
\node [every neuron/.try, neuron1/.try] () at (6,+1.1) {$\Sigma$|R};
\node [every neuron/.try, neuron1/.try] () at (6,+0.2) {$\Sigma$};

\node [right] () at (6.25,+2) {standard scaler};
\node [right] () at (6.25,+1.1) {perceptron + ReLU};
\node [right] () at (6.25,+0.2) {perceptron};

\node [above] at (hidden1-1.north east) {$H_1$};
\node [above] at (hidden1-2.north east) {$H_{n_{1}}$};

\node [above] at (hidden2-1.north east) {$H_1$};
\node [above] at (hidden2-2.north east) {$H_{n_{2}}$};

\foreach \l [count=\i] in {1}
  \draw [->] (output-\i) -- ++(1,0)
    node [above] {{12+log(O/H)}};

\foreach \i in {1,...,4}
  \foreach \j in {1,...,2}
    \draw [->] (input-\i) -- (hidden1-\j);

\foreach \i in {1,...,2}
  \foreach \j in {1,...,2}
    \draw [->] (hidden1-\i) -- (hidden2-\j);

\foreach \i in {1,...,2}
  \foreach \j in {1}
    \draw [->] (hidden2-\i) -- (output-\j);


\node [align=center, above] at (0*2,2) {Input\\layer};
\node [align=center, above] at (1*2,2) {Hidden\\layer(s)};
\node [align=center, above] at (2*2,2) {Output\\layer};

\end{tikzpicture}
\caption{A schematic showing the architecture of the simple feed-forward, fully-connected, multi-layer perception model. The MLP model has three main components. The input layer ingests 16 features (see Table~\ref{tbl_features}) and passes the features through standard scaler functions. The hidden layers comprise of $n_1$ and $n_2$ perceptrons in the first and second layer, restrictively. Each neuron in the hidden layer is a perceptron with an activation function attached. The activation function adopted here is the rectified linear unit (ReLU). The output layer consists of a single perceptron without activation function attached and is used to predict the oxygen abundance. In the main text, models with 1 and 2 hidden layers are tested. For the final model, only 1 hidden layer is adopted. }\label{fig_MLP}
\end{figure*}
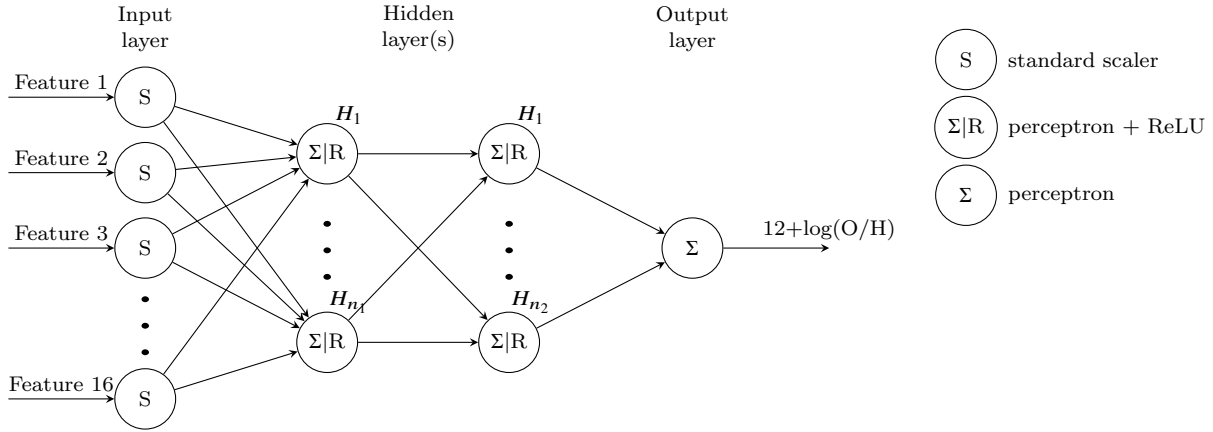


\section{Methods: combining literature auroral line data and a simple artificial neural network}

\subsection{Sample}

The \HII\ region sample is drawn from two sources, containing originally 1009 oxygen abundances estimated based on auroral line(s). A sub-sample of 956 regions forms the final sample for this study. 

The first source is 858 literature \HII\ region data compiled first by \citet{Pilyugin:2012fk} and \citet{Pilyugin:2016aa}. An updated version of this literature compilation is adopted here. The sample contains 797 regions where the [\ion{O}{iii}]$\lambda 4363$ line is detected and 151 regions the [\ion{N}{ii}]$\lambda 5755$ is detected. A small fraction of the sample has both auroral lines 90. The \te\ oxygen abundances are derived following the prescription described in \citet{Pilyugin:2012fk} and assuming the commonly adopted t2 -- t3 scaling relationship \citep{Campbell:1986aa,Garnett:1992aa}. When both auroral lines are detected which yields two estimates of the oxygen abundance, the average is adopted. The second source comes from the on-going CHemical Abundances Of Spirals (CHAOS) survey conducted with the Multi-Object Double Spectrograph (MODS) on the Large Binocular Telescope (LBT). In total, 151 \HII\ regions are drawn the three CHAOS publications, 45 from \citet[CHAOS~\Romannum{1}]{Berg:2015aa}, 29 from \citet[CHAOS~\Romannum{2}]{Croxall:2015aa} and 77 from \citet[CHAOS~\Romannum{3}]{Croxall:2016tg}. The oxygen abundances tabulated in their Tables~5, 5, and 4 are adopted, respectively.

Figure~\ref{fig_sample} presents the sample distributions. The P16 sample contains mostly low-metallicity regions that produce bright [\ion{O}{iii}]$\lambda 4363$. The CHAOS survey fills in the high-metallicity regime by virtue of the sensitivity of MODS and its capability to probe multiple auroral lines simultaneously ([\ion{O}{iii}], [\ion{N}{ii}], [\ion{S}{iii}], [\ion{O}{ii}] and [\ion{S}{ii}]). 

To arrive at the final, working sample (956 regions), an abundance cutoff of 7.4 is imposed to remove the long low-metallicity tail. This is to avoid the calibration being affected by the sparse sampling in a relatively large metallicity space. Figure~\ref{fig_bpt} shows where the working sample lies in the BPT line ratio space \citep{Baldwin:1981lr}.

\subsubsection{Bootstrapping}
The working sample does not uniformly cover the metallicity range of [7.4, 8.9], despite removing the low metallicity tail and including the high-metallicity CHAOS regions. To avoid the fitting being biased by the non-uniform distribution, I bootstrap the working sample to generate uniformly distributed samples.

The (non-uniform) working sample is first sorted and grouped in bins of approximately 30 regions. In each bin, new samples are drawn randomly (with replacement) and added to the bin such that the new sample number is proportional to the width of the bin. This re-sampling is done for every bin, except for the reference bin where the bin width is the narrowest. The reference bin defines the targeted bin-width to sample number ratio. The bootstrapped samples, each containing 2516 regions, are adopted to fit neural network models and quantify their performance. Performing analysis on the bootstrapped samples is equivalent to imposing higher weights on regions having high or low metallicities.

\subsection{Multi-layer perceptron}

I build a simple, feed-forward artificial neural network using the multi-layer perceptron (MLP) framework. A schematic of the network architecture is shown in Figure~\ref{fig_MLP}. The MLP is constructed using \textsc{scikit-learn}\footnote{\url{http://scikit-learn.org/}} \citep{scikit-learn} version v0.19.2 in \textsc{python3}. The fully-connected MLP has three components: input layer, hidden layer(s), and output layer. 

The input layer ingests ``features'', i.e.~logarithmic line ratios, from the training data. The 16 features (based on strong forbidden lines) are listed in Table~\ref{tbl_features}. Here, all the line fluxes are normalised to H$\beta$ and the logarithmic line ratios are adopted as the input features. To optimise the MLP performance, the 16 inputs are first passed through standard scaler functions such that the means are zero and the standard deviations are unity.

The 16 features are selected by taking possible arithmetic combinations of line ratios, i.e.~single line, ratio of line pairs and sum of line pairs. Some of the features are line ratios well-known to be sensitive to metallicity. For example, the [\ion{N}{ii}], [\ion{N}{ii}]/[\ion{O}{ii}], and [\ion{N}{ii}]/[\ion{O}{iii}] are popular diagnostics \citep[e.g.][]{Kewley:2002fj,Pettini:2004lr}. The combination of [\ion{S}{ii}]/[\ion{N}{ii}] and [\ion{N}{ii}] is also shown to be a robust metallicity indicator \citep{Dopita:2016fk}.
[\ion{O}{iii}]/[\ion{O}{ii}] is known to be sensitive to both ionisation parameter and metallicity \citep[e.g.][]{Kewley:2002fj}. [\ion{O}{ii}] + [\ion{O}{iii}], or equivalently $\rm R_{23}$, is yet another widely adopted metallicity indicator \citep{Pagel:1979aa}. The rest of the features are not traditionally adopted in metallicity diagnostics. These line ratios could be sensitive to other physical properties of the ISM which may help improve the predication of metallicity. The intention is to test if by non-linearly combining these features using the neural network, a better model could emerge.

\begin{figure*}
\includegraphics[width = \textwidth]{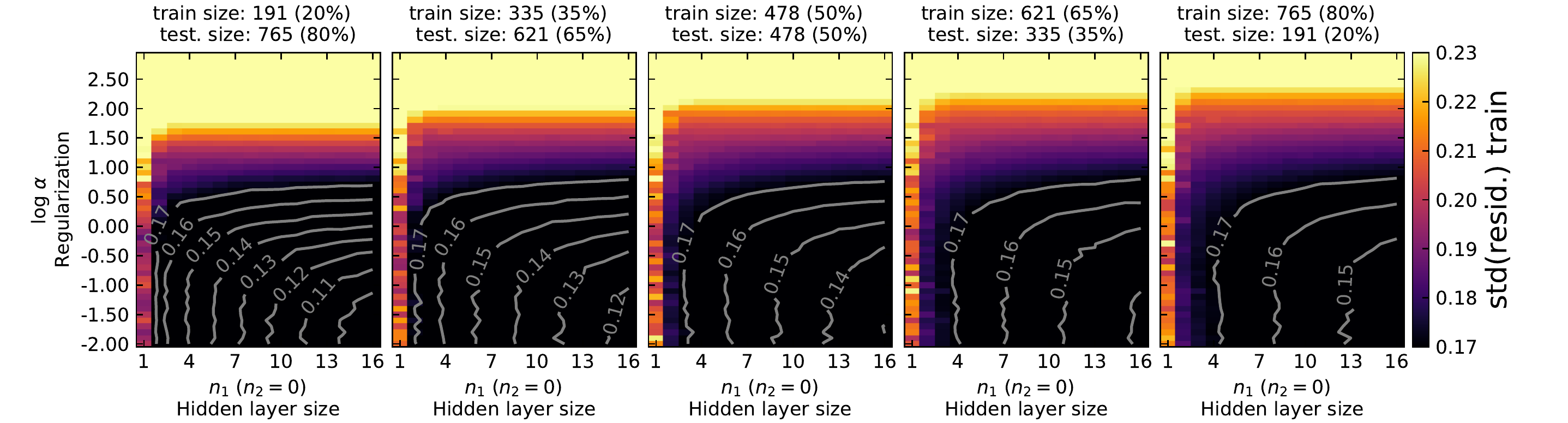}
\caption{In-sample performance for 1 hidden layer models at different model complexities and training sample sizes. The training sample size increases from the left to the right panels. The training and testing sample sizes (and fractions to the working sample) are labeled on the top of each panel. In each panel, different combinations of hyper-parameters $(\log\alpha,n_1)$ correspond to different model complexities. The colour scale indicates the standard deviation of the residual of the training sample. Details are discussed in Section~2.3.}\label{fig_explore_params_2}
\end{figure*}

In the hidden layers, each neuron is a simple perceptron with the output being passed through an activator function. Each perceptron performs a simple calculation 
\begin{equation}
p(x) = \sum_i w_i x_i + b,
\end{equation}
where $w_i$ and $b$ (i.e.~weight and bias) are the model variables and $x_i$ are input from the previous layer. The rectified linear unit is adopted as the activator function that adds nonlinearity to the model. That is:
\begin{equation}
A(x) =  \max(0,x).
\end{equation}
In other words, each neuron in the hidden layers ingests outputs of the previous layer, $x_i$, and performs the calculation
\begin{equation}
A(p(x)) = \max(0,\sum_i w_i x_i + b). 
\end{equation}
Two models are explored below: one hidden layer with $n_1$ neurons and two hidden layers with ($n_1$,$n_2$) neurons. The final, best model has only one hidden layer. 

The output layer comprises of one single perceptron with no activator function attached. The perceptron in the output layer is used to predict the targeted unknown quantity, i.e.~oxygen abundance.

The MLP framework described above is controlled by two hyper-parameters. The first parameter is the number of neurons in the hidden layers, $n_1$ and $n_2$. A larger number of neurons in the hidden layers allows the model a higher degree of complexity. Since complex models are prone to over-fitting the data, a second parameter $\alpha$, the L2 penalty parameter (for regularisation), is introduced. At a fixed ($n_1$,$n_2$), $\alpha$ of zero imposes no prior on all the possible models under the MLP configuration. A non-zero, positive $\alpha$ penalises the chi-square by the sum of the weight squares (multiplied by $\alpha$), effectively guiding the fitting towards models with neuron weights closer to zero.

\subsection{Train and test the neural network}
Training the MLP network was done using the standard solver implemented in \textsc{scikit-learn}. The weights of the perceptrons are initialised randomly, and then a limited-memory version of the Broyden-Fletcher-Goldfarb-Shanno algorithm was adopted to minimise the loss function. The minimisation terminated when the the loss function did not improve by $10^{-4}$ for two consecutive iterations. Since the training sample is moderate in size, all the training sample was used simultaneously without performing any stochastic minimisation using mini-batches.

\subsubsection{Hyper-parameter tuning}
In general, hyper-parameter tuning is required to avoid over-training a neural network, i.e.~using a model too complex for the available data, resulting in the model to ``remember the answers by-heart''. This problem is usually avoided by withholding a part of the data to test the model performance, and tuning the complexity of the model until the model performs equally well on data that it has and has not seen (i.e.~training and testing samples). Below, I systematically investigate how the two hyper-parameters affect the network performance. This investigation will be done at different sample sizes to further understand whether more complex models can be adopted with increasing sample sizes. 

For a give combination of hyper-parameters and train-sample size, a training sample is randomly drawn from the working sample and the remaining data are used for testing. Both the training and testing samples are bootstrapped (separately) to avoid the non-uniformity discussed in Section~2.1.1. After the model is trained, the standard deviation of the residuals, i.e.~predicted versus \te\ metallicities, is adopted to quantify the model performance. Both the residuals of the training and testing sets are calculated to quantify the in- and out-of-sample performance, respectively. 

To avoid my analysis being biased by the random components described above, i.e.~the random weight initialisation, the random drawing of train/test samples and the bootstrap re-sampling, the sample drawing and fitting are repeated 100 times with different random states. This yields 100 performance pairs at each hyper-parameter grid point. The mean values are presented henceforth to understand the appropriate hyper-parameters and MLP performance.

\begin{table}
\caption{Input features ($N=16$)}
\label{tbl_features}
\begin{tabular}{ll}
\hline
Type &  Line ratio \\
\hline
Single line & [\ion{O}{ii}], [\ion{O}{iii}], [\ion{N}{ii}], [\ion{S}{ii}] \\
Ratio of line pairs & [\ion{O}{iii}]/[\ion{O}{ii}], [\ion{N}{ii}]/[\ion{O}{ii}], [\ion{S}{ii}]/[\ion{O}{ii}], \\
& [\ion{N}{ii}]/[\ion{O}{iii}], [\ion{S}{ii}]/[\ion{O}{iii}], [\ion{S}{ii}]/[\ion{N}{ii}]\\
Sum of line pairs & 
([\ion{O}{ii}]+[\ion{O}{iii}]), 
([\ion{O}{ii}]+[\ion{N}{ii}]), \\
& ([\ion{O}{ii}]+ [\ion{S}{ii}]), 
([\ion{O}{iii}]+[\ion{N}{ii}]), \\
& ([\ion{O}{iii}]+[\ion{S}{ii}]), 
([\ion{N}{ii}]+[\ion{S}{ii}])\\
\hline
\multicolumn{2}{l}{1. All lines are normalized to H$\beta=1$}\\
\multicolumn{2}{l}{2. The input features are the {\em logarithmic} of the line ratios}\\
\end{tabular}

\end{table}


\subsubsection{One hidden layer}

\begin{figure*}
\includegraphics[width = \textwidth]{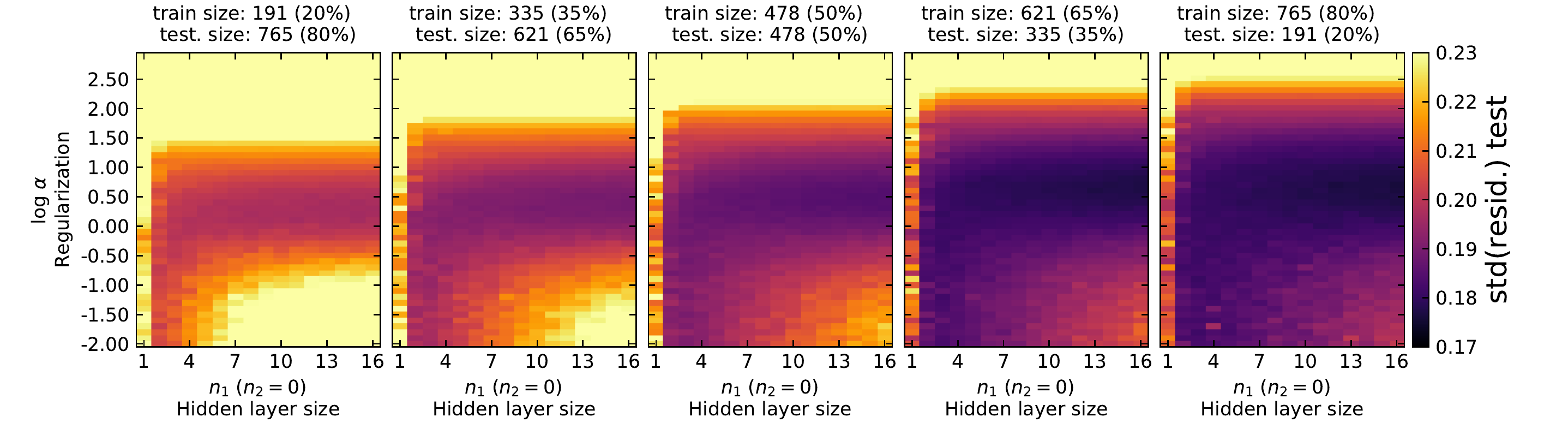}
\caption{Out-of-sample performance for different model complexities and training sample sizes. Same as Figure~\ref{fig_explore_params_2} but for the testing sample. Details are described in Section~2.3.}\label{fig_explore_params_1}
\end{figure*}

I start by training a network with only one hidden layer and $n_1$ neurons. I explore a grid of hyper-parameters with $n_1$ from 1 to 16 neurons with a step of 1, and $\log\alpha$ from $-2.0$ to $+2.9$ with a step of 0.1. I explore the parameter grids using five different training sample sizes, from 20\% (191) to 80\% (765) of the working sample in a step of 15\% (143). In Figures~\ref{fig_explore_params_2} and \ref{fig_explore_params_1}, I present the in- and out-of-sample performance at different sample sizes and model complexity, respectively. The sample size increases from the left to the right panels. Within each panel, the model complexity increases with increasing $n_{1}$ and decreasing $\alpha$.

In Figure~\ref{fig_explore_params_2}, for a given sample size the in-sample performance in general improves with increasing model complexity. For a given model complexity, more training data leads to worse performance. That is, when the sample size becomes larger, a more complex model is required to improve the in-sample performance. 

Figure~\ref{fig_explore_params_1} shows the out-of-sample performance. In general, models that are either too simple or too complex (i.e.~close to the edge of the parameter space explored) have poor performance. At each sample size, there is a family of models that can perform equally well.  With increasing sample size, the achievable best-performance improves. That is, with more \HII\ regions complex models can be developed, which achieve better out-of-sample performance. 

To avoid over-fitting the data, the hyper-parameters need to be chosen to deliver comparable in-and and out-of-sample performance with the latter being as good as possible. This can be done by comparing Figures~\ref{fig_explore_params_2} and \ref{fig_explore_params_1}. At a given sample size, the appropriate model complexities can be achieved by different combinations of $(\log\alpha,n_1)$, networks with fewer neurons and less regularisation (smaller $\alpha$ and $n_1$) or with more neurons and more regularisation (larger $\alpha$ and $n_1$). The parameter $\log\alpha$ of approximately $1.0$ appears to avoid over-training over a wide range of $n_1$.

Finally, it is worth noting that the $n_1=1$ models are not optimal in all five sample sizes. Even with little regularisation, the single neuron model is still too simple to fully extract information from the data. The performance difference between $n_1=1$ and $n_1>1$ models is increasingly more obvious with increasing sample size. The single neuron model describes essentially linear combinations of the 16 logarithmic line ratios. Some existing strong line calibrations only fit linear combinations of logarithmic line ratios to the data, e.g.~using the N2 or O3N2 indexes. These calibrations are similar to the single neuron models. This suggests that properly trained MLP models with more than one neuron will out-perform these literature calibrations. In other words, given the available literature samples complex models utilising multiple line ratios could better calibrate strong line metallicity than those using only one or two line ratio pairs.

\subsubsection{Two hidden layers}
\begin{figure}
\includegraphics[width = \columnwidth]{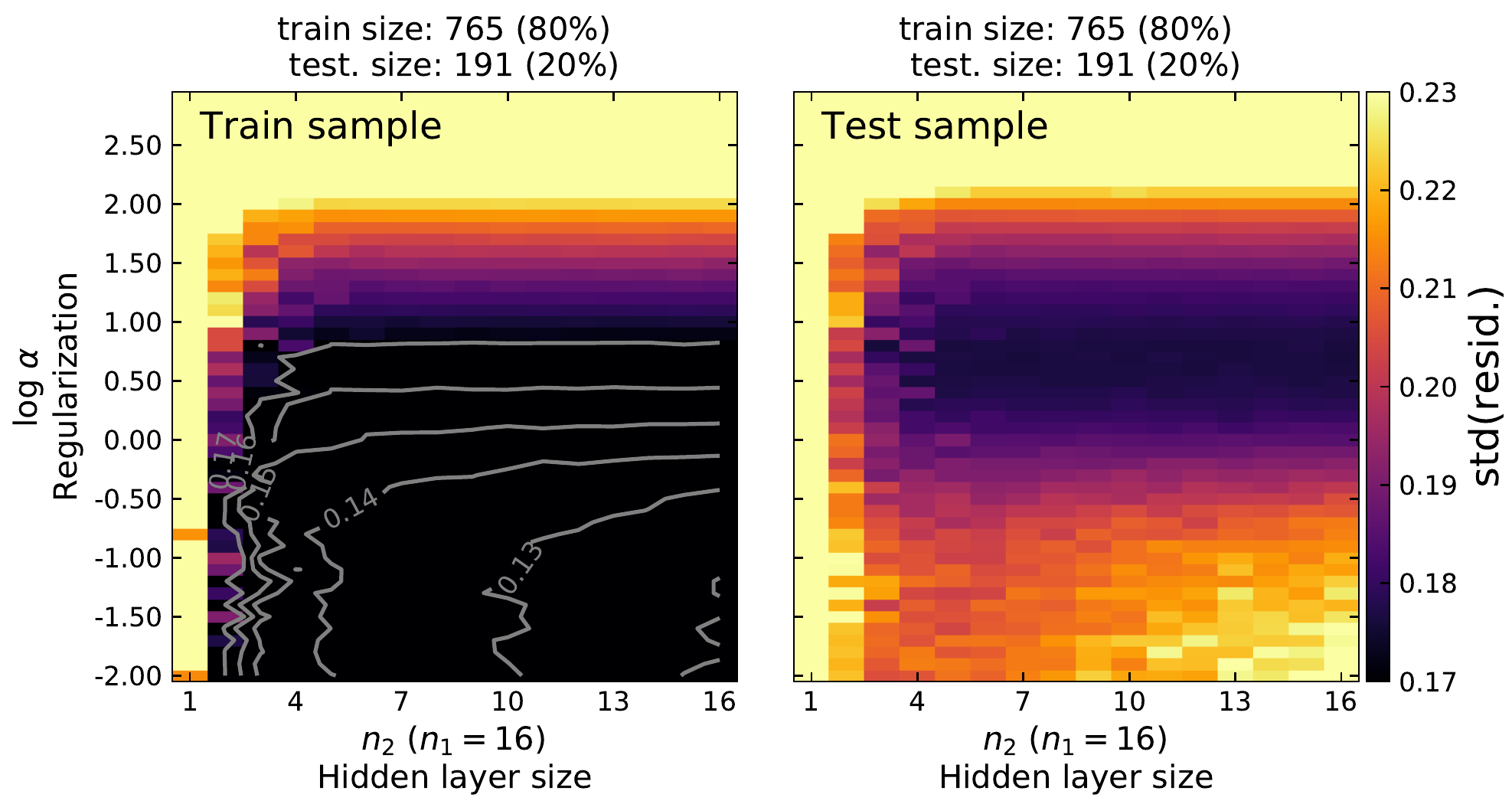}
\caption{In- and out-of-sample performance for 2 hidden layer models at different model complexities (left and right panels, respectively). The first hidden layer has 16 neurons. The second layer has up to 16 neurons (see Figure~\ref{fig_MLP}). }\label{fig_explore_params_4}
\end{figure}

The performance of a single hidden-layer network saturates at large $n_1$. It is worth exploring whether including a second hidden-layer could further increase the network performance. Here, I fix $n_1$ to 16 and systematically explore the network performance at different $n_2$ and $\alpha$. This exploration is done only for the 80\%/20\% train/test split. As before, 100 models with different random states are trained for each grid point. In Figure~\ref{fig_explore_params_4}, I show the mean performance of the 100 realisations. 

Figure~\ref{fig_explore_params_4} demonstrates that with the inclusion of a second hidden layer, the in-sample performance improves further, from 0.15~dex in the right panel of Figure~\ref{fig_explore_params_2} to 0.12~dex. However, the out-of-sample performance remains virtually the same. This suggests that at the current sample size, an additional hidden layer is not critical for fitting the data.

\subsubsection{Best MLP model}

I have demonstrated how the hyper-parameters and sample size affect the choice of model complexity, and showed that one hidden layer is sufficient for the current sample size. Here, I train the final MLP model using all the \HII\ region sample, without withholding any data for testing. I adopt the hyper-parameters $(\log\alpha, n_1, n_2)=(1.2, 10, 0)$. This parameter combination delivers in- and out-of-sample performance of approximately 0.18 and 0.18, respectively, for the 80\%/20\% train/test split (right panels in Figures~\ref{fig_explore_params_2} and \ref{fig_explore_params_1}). Given more training data for the final model, these hyper-parameters should deliver an appropriate model complexity that does not over-fit the data. As before, I train 100 models with different random states to avoid being biased by the random components in the fitting process. Henceforth, I will report the median value of the predictions from the 100 models.

\begin{figure}
\centering
\includegraphics[width = 0.8\columnwidth]{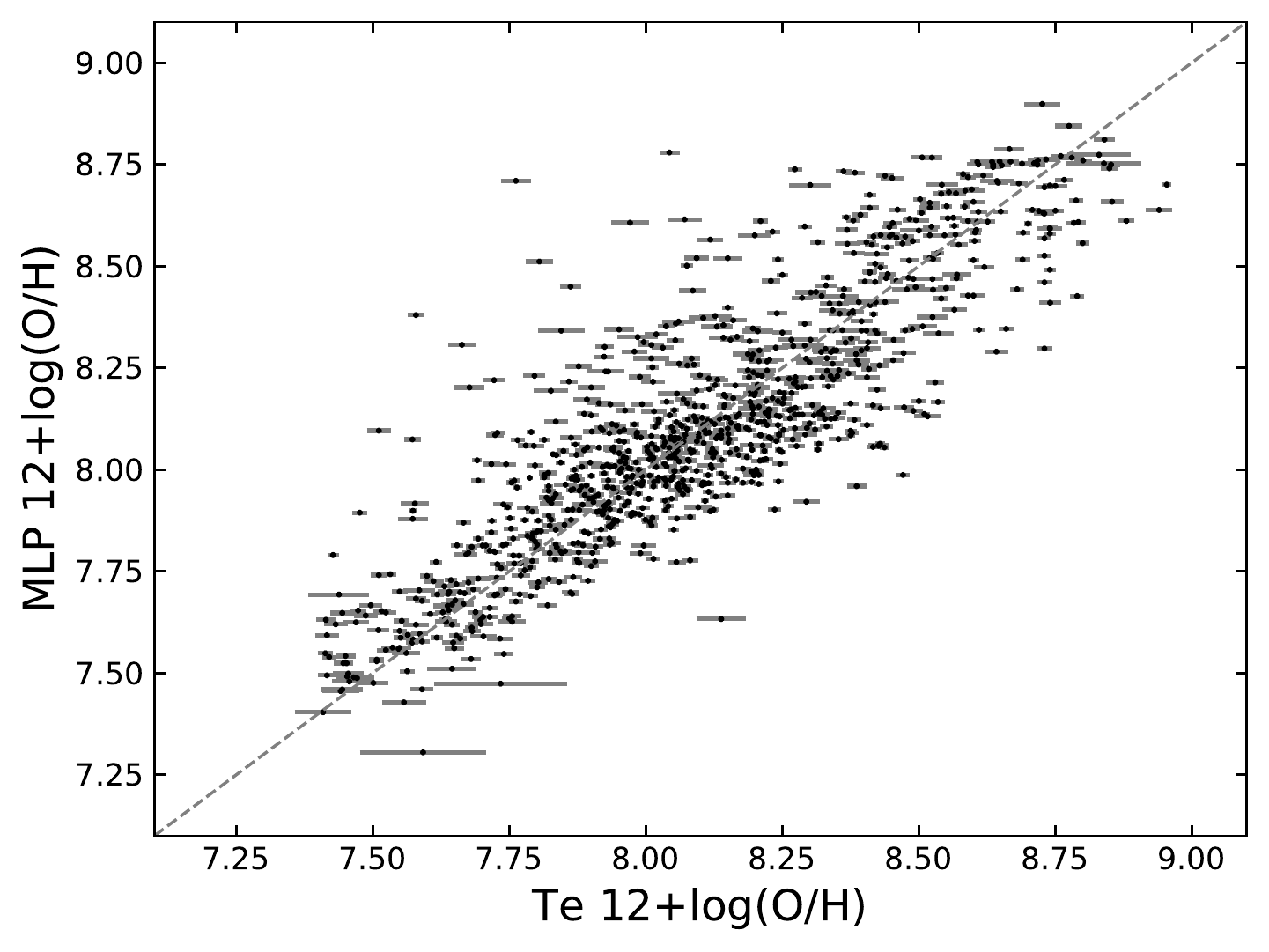}
\caption{Comparison between oxygen abundances from the best MLP model (Section~2.3.4) and the \te\ method. The data points are the working sample described in Section~2.1. The error bars are the standard deviations of the 100 predictions from the 100 model realizations. The dashed line is the one-to-one line.}\label{fig_compare_MLP}
\end{figure}

In Figure~\ref{fig_compare_MLP}, I compare input \te\ metallicities with those predicted by the MLP model from strong lines. The error bars show the 1$\sigma$ standard deviations of the 100 predictions. In general, the spread in metallicity due to the random components in the fitting process is much smaller than the dispersion of the calibration.  Figure~\ref{fig_compare_MLP} suggests that the best MLP model performs reasonably well over a wide range of metallicity.

\subsubsection{Feature sensitivity}

\begin{figure}
\centering
\includegraphics[width = \columnwidth]{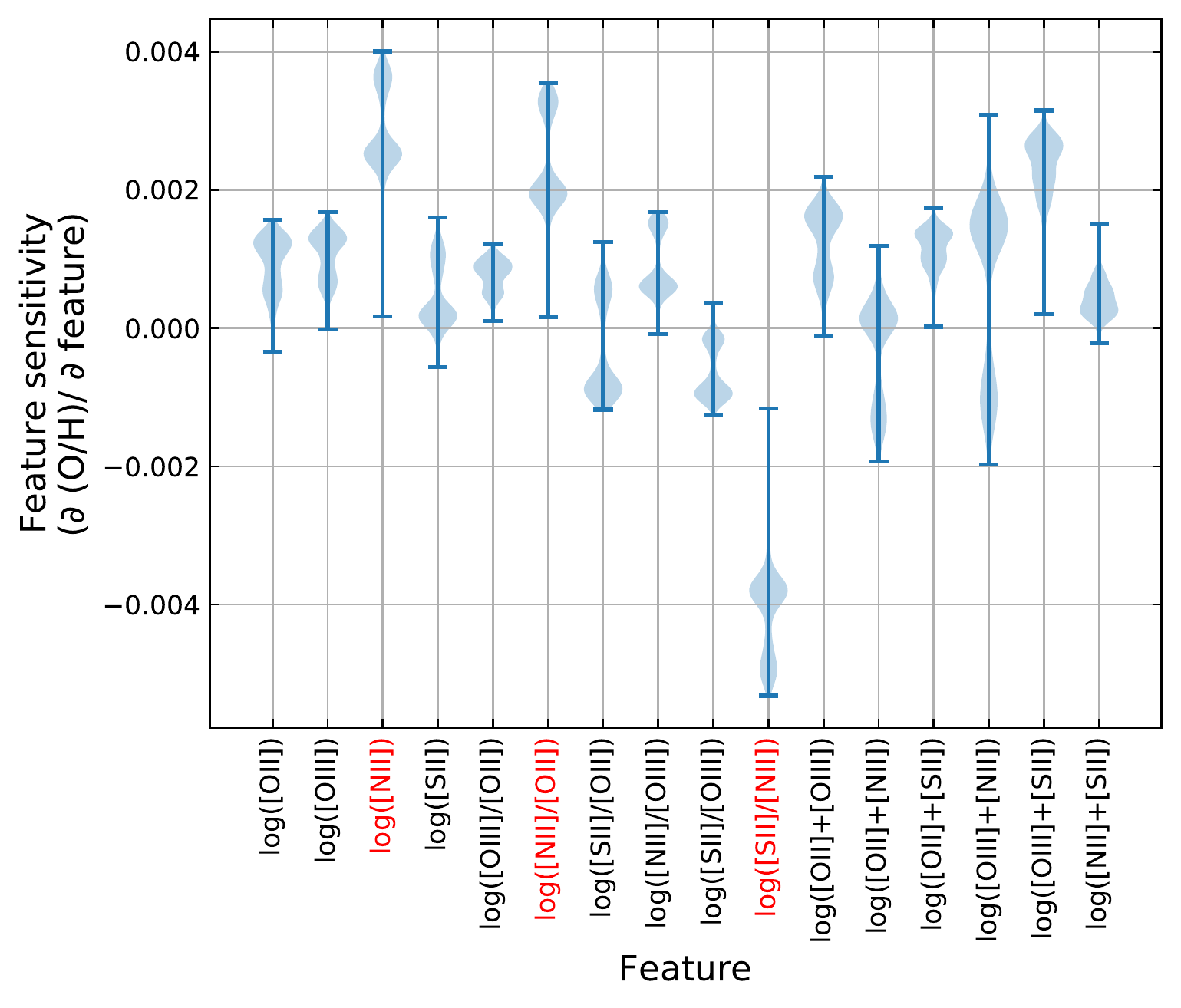}
\caption{
Feature sensitivity of the best MLP model, as measured by perturbing one feature at a time and recording the response of the output oxygen abundance. The response of the working sample to different feature perturbations is shown as the shaded vertical histograms. The top three important features are labeled in red (i.e.~deviate most from zero). }\label{fig_feature_sensitivity}
\end{figure}

Extracting physical interpretation from neuron weights and biases is known to be difficult. However, it is worth investigating how sensitive different features are to the output to understand their importance to the network. The sensitivity analysis can be achieved by computing the partial derivative of the output with respect to the input features. 

Using the best MLP model, I perturb one feature at a time while keeping the other features unchanged and compute the changes in the output metallicity. The small perturbation is set to 3\% of the standard deviation of the input feature of the working sample. The response of the model to the perturbations is calculated for the entire working sample. Figure~\ref{fig_feature_sensitivity} presents histograms of the resulting changes in the output metallicities in the form of a ``violin'' plot. The top three features that the network output is most sensitive to are [\ion{S}{ii}]/[\ion{N}{ii}], [\ion{N}{ii}]/[\ion{O}{ii}], and [\ion{N}{ii}] (deviate most from zero). Interestingly, photoionisation models have demonstrated that the [\ion{N}{ii}]/[\ion{O}{ii}] ratio is a good metallicity diagnostic (at above 0.2 solar) due to its insensitivity to ionisation parameter \citep{Kewley:2002fj,Dopita:2013qy}. The combination of [\ion{N}{ii}]/[\ion{S}{ii}] and [\ion{N}{ii}]/H$\alpha$ is also suggested by \citet{Dopita:2016fk} to be an excellent diagnostic that is insensitive to ISM pressure and ionisation parameter. The neural network appears to rely more heavily on features that are also important in diagnostics developed from photoionisation models, despite that this information is not given a priori. This demonstrates the ability of the neural network to directly ``learn'' from the data.

\begin{figure}
\includegraphics[width = \columnwidth]{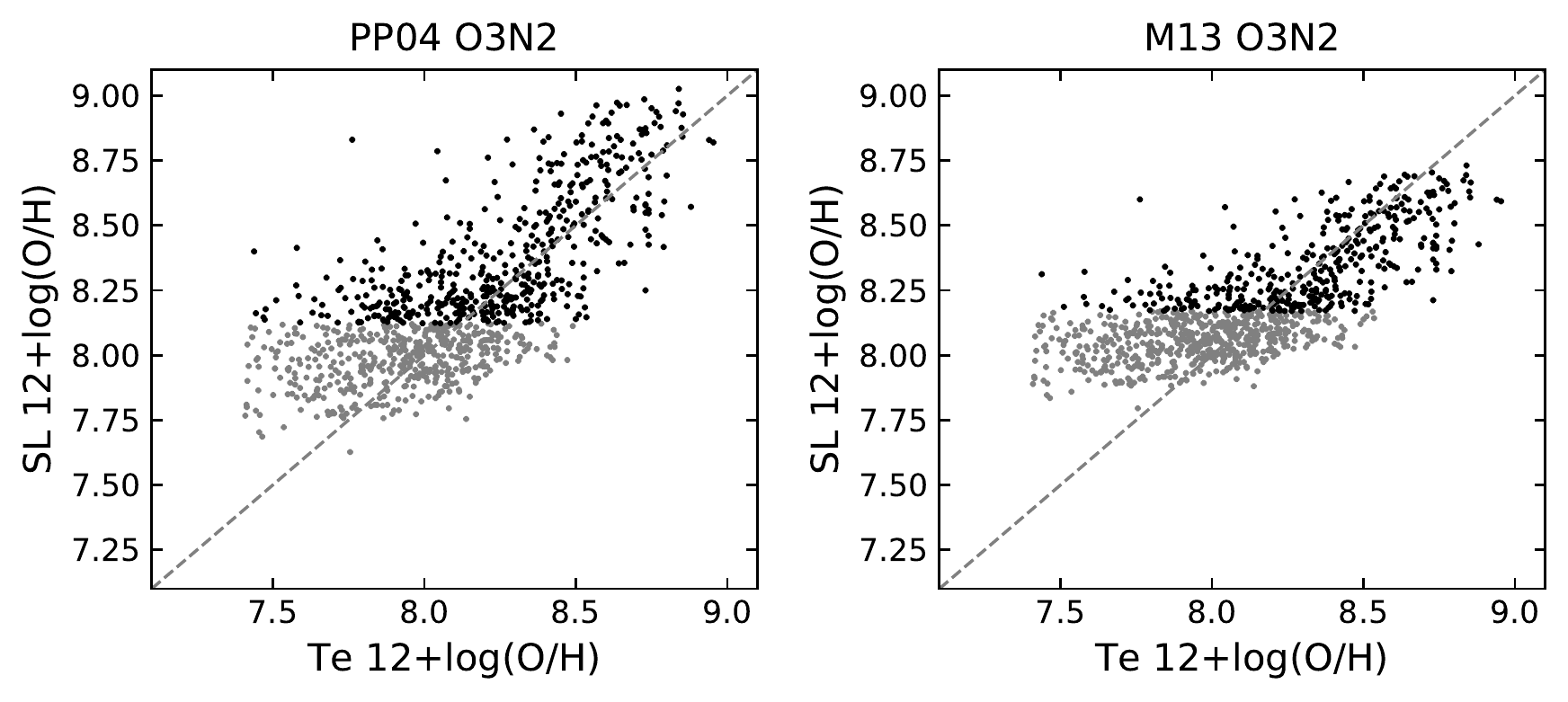}
\includegraphics[width = \columnwidth]{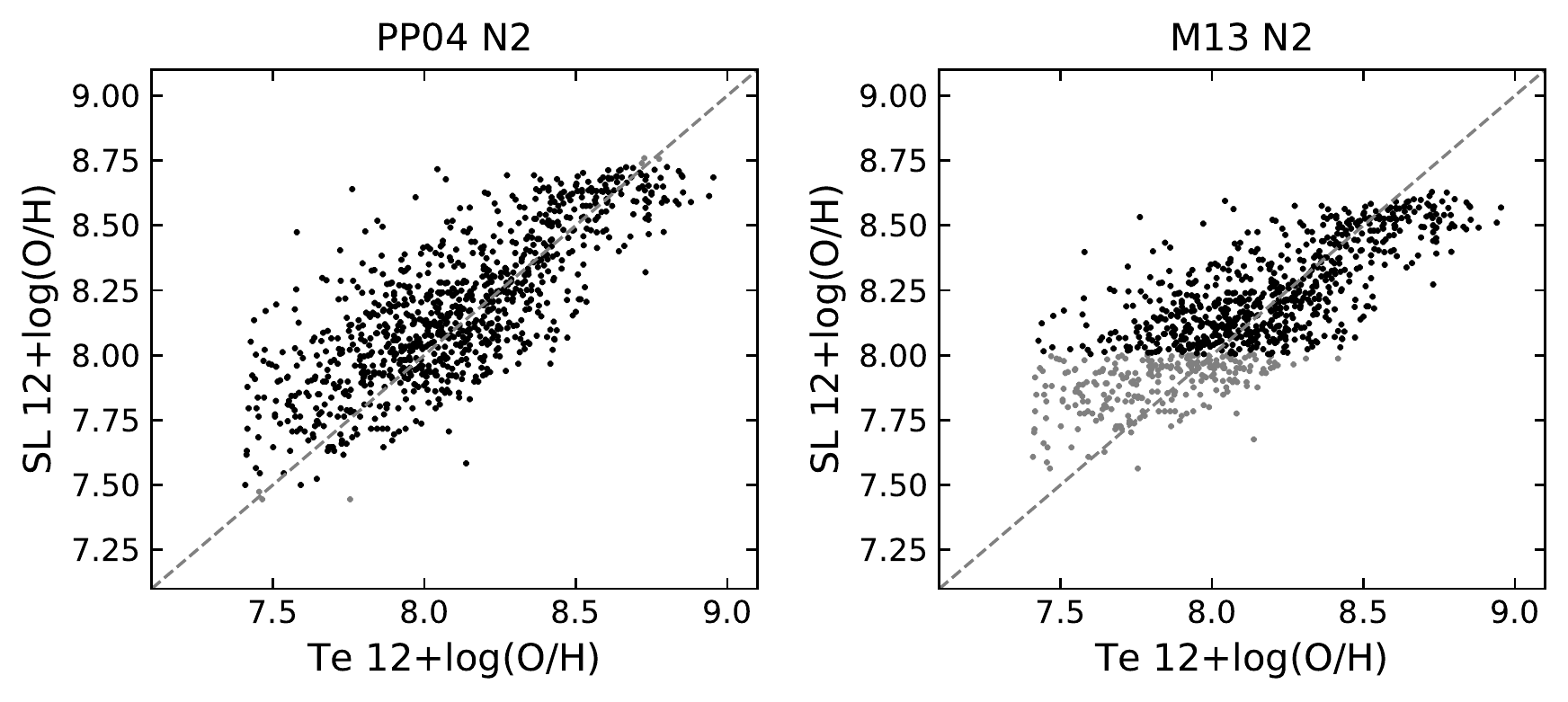}
\includegraphics[width = \columnwidth]{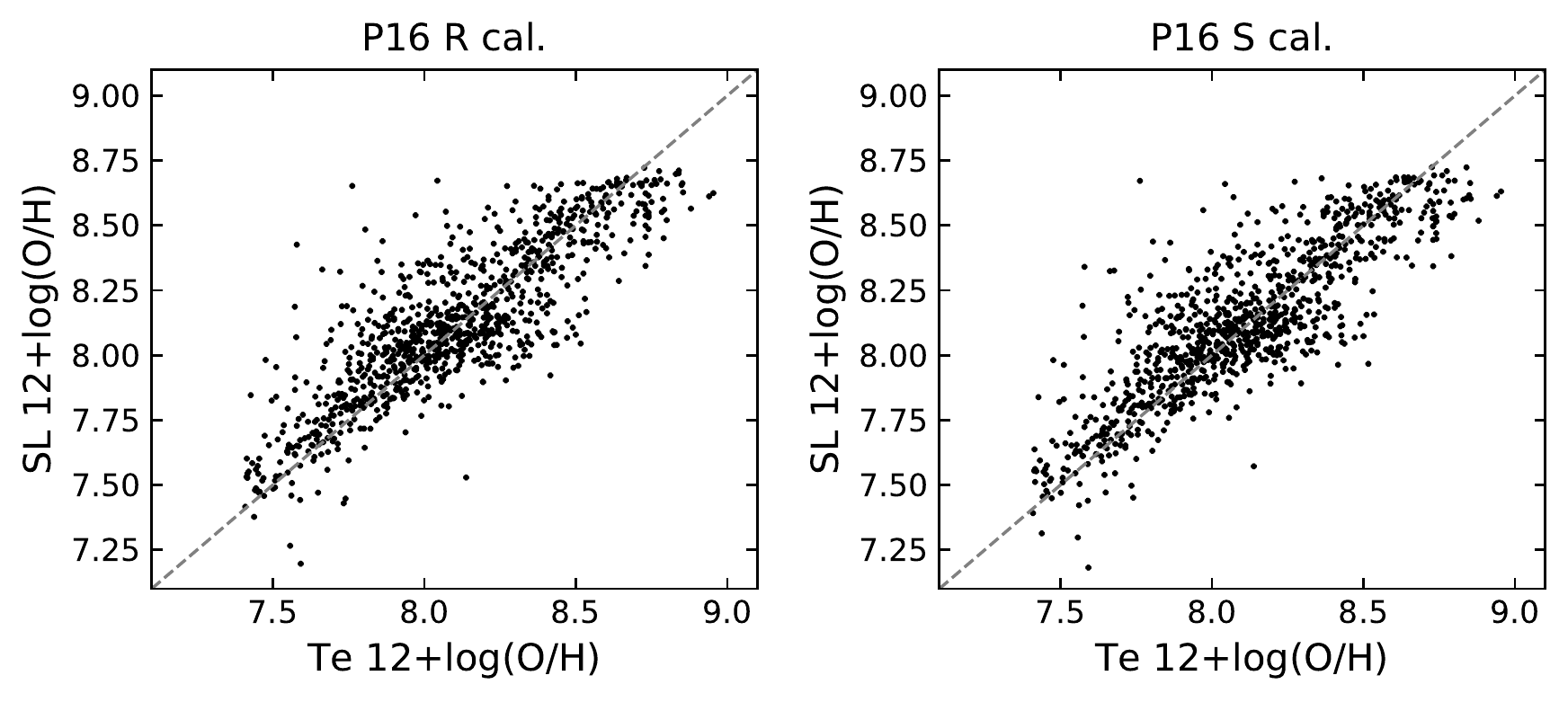}
\caption{Comparison between various strong-line  (y-axes) and \te\ oxygen abundances (x-axes). The dashed lines represent the one-to-one line. }\label{fig_compare_various}
\end{figure}

\section{Results}

\begin{figure*}
\includegraphics[width = 0.8 \textwidth]{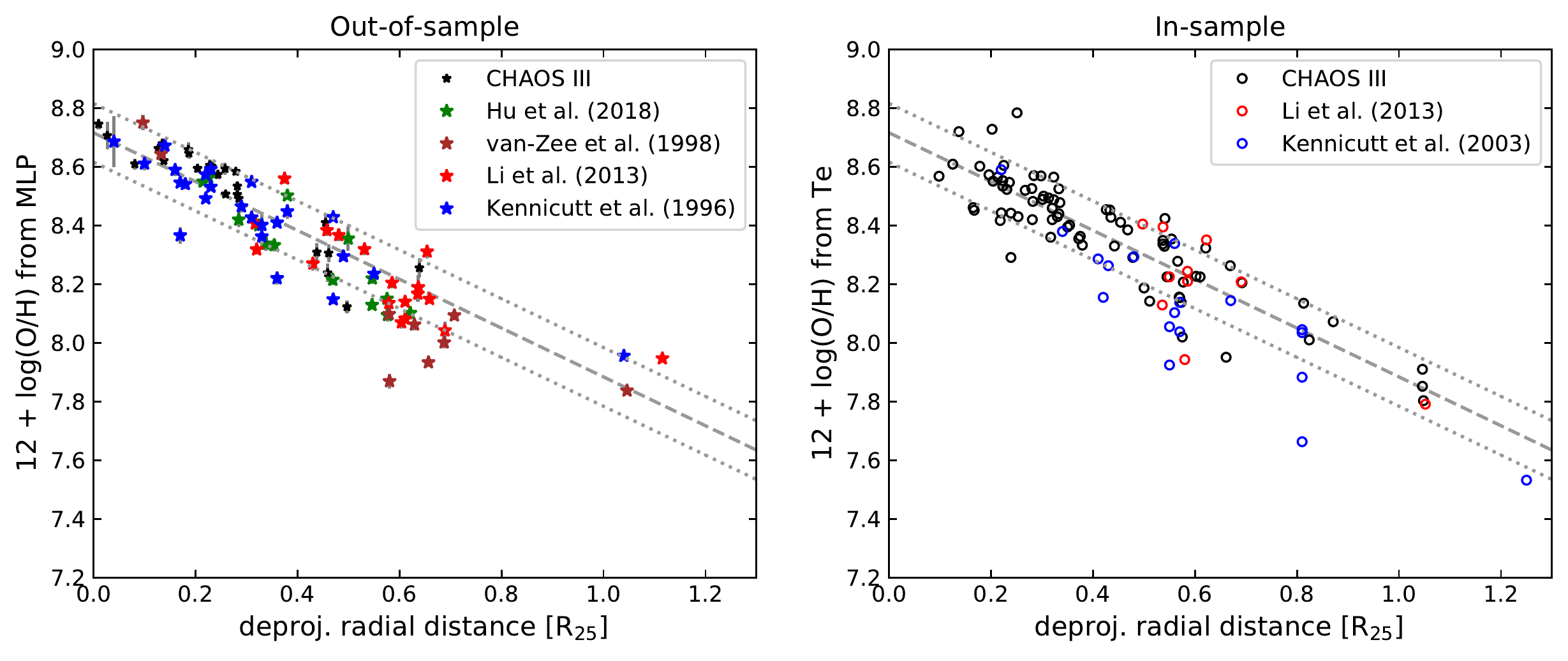}
\caption{ Metallicity gradient of M101. 
{\it Left:} the strong-line metallicities are derived using the MLP model and \HII\ region data that the model has not seen before (out-of-sample). {\it Right:} \te\ metallicities of \HII\ regions in the training sample (in-sample). The dashed lines indicate the best-fit gradient reported by \citep{Croxall:2016tg}. The dotted lines are $\pm0.1$~dex of the best fit. }\label{fig_m101}
\end{figure*}

\subsection{Compare with other calibrations}

It is worth comparing the MLP performance with other literature strong line calibrations to examine the putative improvement. Below, I apply different strong line diagnostics to the working \HII\ region sample and quantify their performance using the metallicity residuals, i.e.~strong-line versus \te\ metallicity.

In Figure~\ref{fig_compare_various}, I compare the strong line diagnostics using the O3N2 and N2 indexes calibrated by \citet{Pettini:2004lr}\footnote{For the N2 calibration, the linear fit is adopted.} and \citet{Marino:2013vn}. The \HII\ regions with indexes beyond the diagnostic limits are shown in grey. The calibrations provide 1D linear functions describing simple relationships between the input indexes and metallicity. Although the O3N2 and N2 indexes are not explicitly included in my feature list (Table~\ref{tbl_features}), equivalent line ratios, [\ion{N}{ii}]/[\ion{O}{iii}] and [\ion{N}{ii}]/H$\beta$, are a part of the input features. In Figure~\ref{fig_compare_various}, I also compare the R and S calibrations developed by P16. These two calibrations each ingests three forbidden to Balmer line ratios. The R calibration uses [\ion{O}{ii}], [\ion{O}{iii}], and [\ion{N}{ii}], and the S calibration [\ion{O}{ii}], [\ion{S}{ii}], and [\ion{N}{ii}]. These calibrations describe more complex 3D mappings from line ratios to metallicity. 

In Tables~\ref{tbl_performance_1} and \ref{tbl_performance_2}, I quantify the residual distributions of the different calibrations. Table~\ref{tbl_performance_1} tabulates the median, mean, and standard deviation derived from the entire sample. Since a large fraction of the sample was adopted to construct the R.- and S.- calibrations, a second comparison in Table~\ref{tbl_performance_2} is done using only the CHAOS sample that has on average a higher metallicity than the P16 sample (see Figure~\ref{fig_sample}). Even with the very simple architecture (Figure~\ref{fig_MLP}) and virtually picking the input features blindly, the MLP out-performs almost all the calibrations with its smallest median, mean, and standard deviation (Table~\ref{tbl_performance_1}). In Table~\ref{tbl_performance_2}, the MLP standard deviation is slightly larger than the R. calibration but its mean and median are smaller. The improvement of MLP over other literature calibrations is somehow marginal (a few to 30 percent in dispersion), however, the prediction accuracy of the ANN method increases with increasing sample size, as demonstrated in Figure~\ref{fig_explore_params_1}. It is likely that when a larger training sample is available, the prediction accuracy can be improved further (see Section~4).

Below, I will apply the MLP calibration to \HII\ region and galaxy integrated spectra to reproduce metallicity gradient and the mass-metallicity relationship (MZR), two of the most fundamental chemical characteristics of star-forming galaxies.

\begin{table}
\caption{Performance comparison (entire sample)}
\label{tbl_performance_1}
\begin{tabular}{lrrr}
\hline
Method & Median & Mean & Standard deviation \\
\hline
MLP (this work) & $ 0.005$ &  $ 0.012$ &  $0.168$ \\
O3N2 (PP04)     & $ 0.111$ &  $ 0.123$ &  $0.228$ \\
O3N2 (M13)      & $-0.011$ &  $ 0.022$ &  $0.222$ \\
N2 (PP04)       & $ 0.063$ &  $ 0.073$ &  $0.206$ \\
N2 (M13)        & $ 0.031$ &  $ 0.048$ &  $0.210$ \\
R cal. (P16)    & $ 0.020$ &  $ 0.023$ &  $0.174$ \\
S cal. (P16)    & $ 0.017$ &  $ 0.023$ &  $0.174$ \\
\hline
\end{tabular}
\end{table}

\begin{table}
\caption{Performance comparison (CHAOS sample)}
\label{tbl_performance_2}
\begin{tabular}{lrrr}
\hline
Method & Median & Mean & Standard deviation \\
\hline
MLP (this work) & $ -0.009$ &  $-0.011$ &   $0.138$  \\
O3N2 (PP04)     & $  0.065$ &  $ 0.058$ &   $0.185$   \\
O3N2 (M13)      & $ -0.089$ &  $-0.091$ &   $0.159$  \\
N2 (PP04)       & $  0.024$ &  $ 0.011$ &   $0.149$   \\
N2 (M13)        & $ -0.067$ &  $-0.072$ &   $0.150$  \\ 
R cal. (P16)    & $ -0.036$ &  $-0.041$ &   $0.133$   \\
S cal. (P16)    & $ -0.026$ &  $-0.043$ &   $0.139$   \\
\hline
\end{tabular}
\end{table}

\subsection{Abundance gradients in nearby galaxies}

I apply the MLP calibration to a sample of \HII\ regions in M101. The M101 galaxy is chosen because of its rich literature \HII\ region data and its well-behaved radial gradient that covers a wide range of metallicity. The MLP calibration is applied to only regions that the model has not seen before. Theses include 25, 9, 18, and 24 regions from \citet{Croxall:2016tg}, \citet{van-Zee:1998yq}, \citet{Li:2013nx} and \citet{Kennicutt:1996yf}, respectively. The auroral lines were undetected in these regions. An additional 15 regions from the recent work by \citet[][their Table~1]{Hu:2018aa} is also included. 

In Figure~\ref{fig_m101}, I show the metallicity gradient from the MLP calibration in the left panel. For reference, I show the \te\ metallicities from \citet{Croxall:2016tg}, \citet{Li:2013nx} and \citet{Kennicutt:2003qd} in the right panel. These \HII\ regions are in the training sample. In both panels, the grey dashed lines indicate the best-fit gradient from \citet{Croxall:2016tg}. Figure~\ref{fig_m101} demonstrates that the MLP calibration produces a reasonable metallicity gradient and comparable scatter, compared with the \te\ method.

\begin{figure}
\includegraphics[width = \columnwidth]{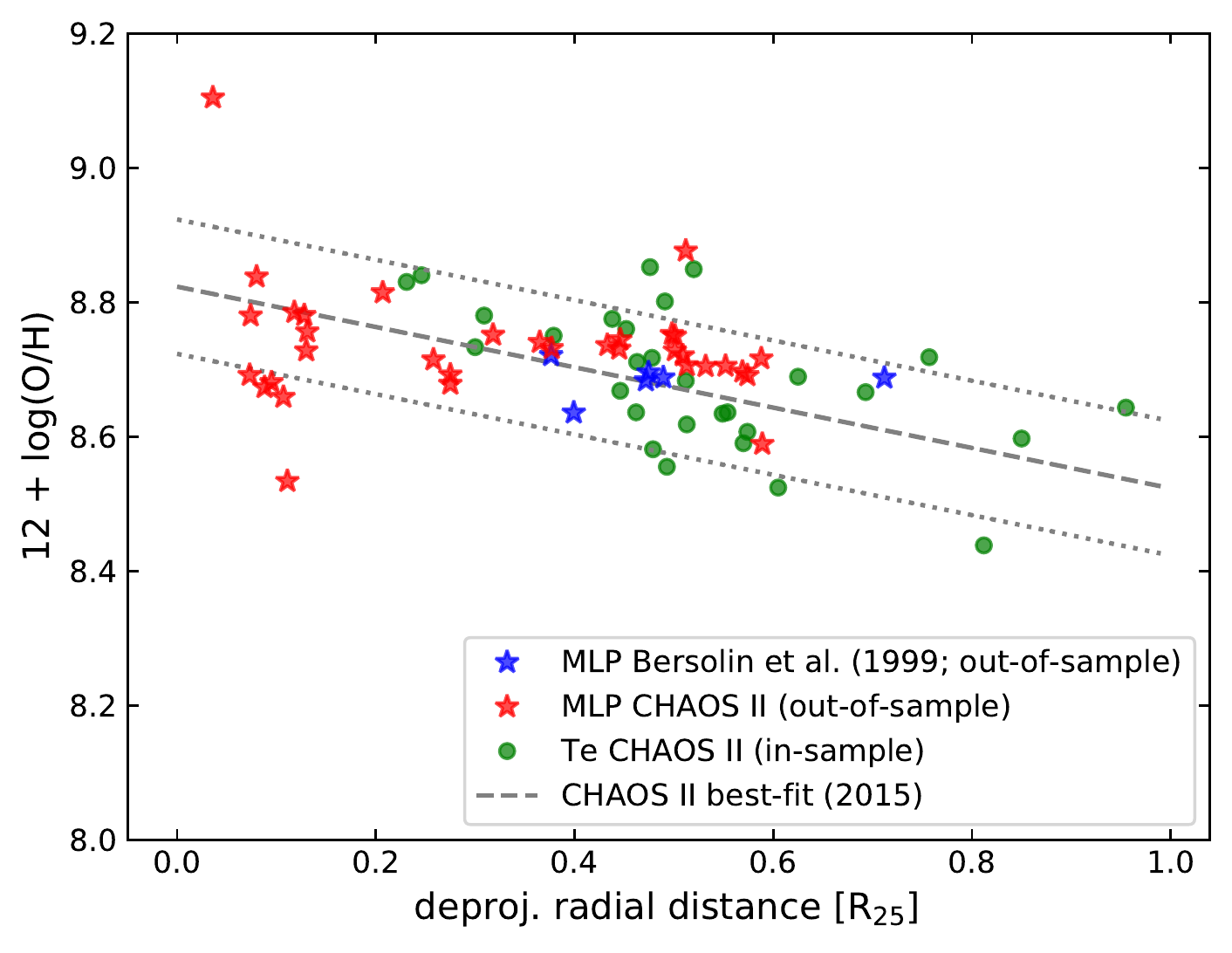}
\caption{Metallicity gradient of M51. The star symbols indicate metallicities derived using MLP for the out-of-sample regions. The green circles represent \te\ metallicities from \citet{Croxall:2015aa}. The dashed lines indicate the best-fit gradient reported by \citep{Croxall:2015aa}, based on the green circles. The dotted lines are $\pm0.1$~dex of the best fit.}\label{fig_m51}
\end{figure}

\begin{figure*}
\includegraphics[width = 0.8\textwidth]{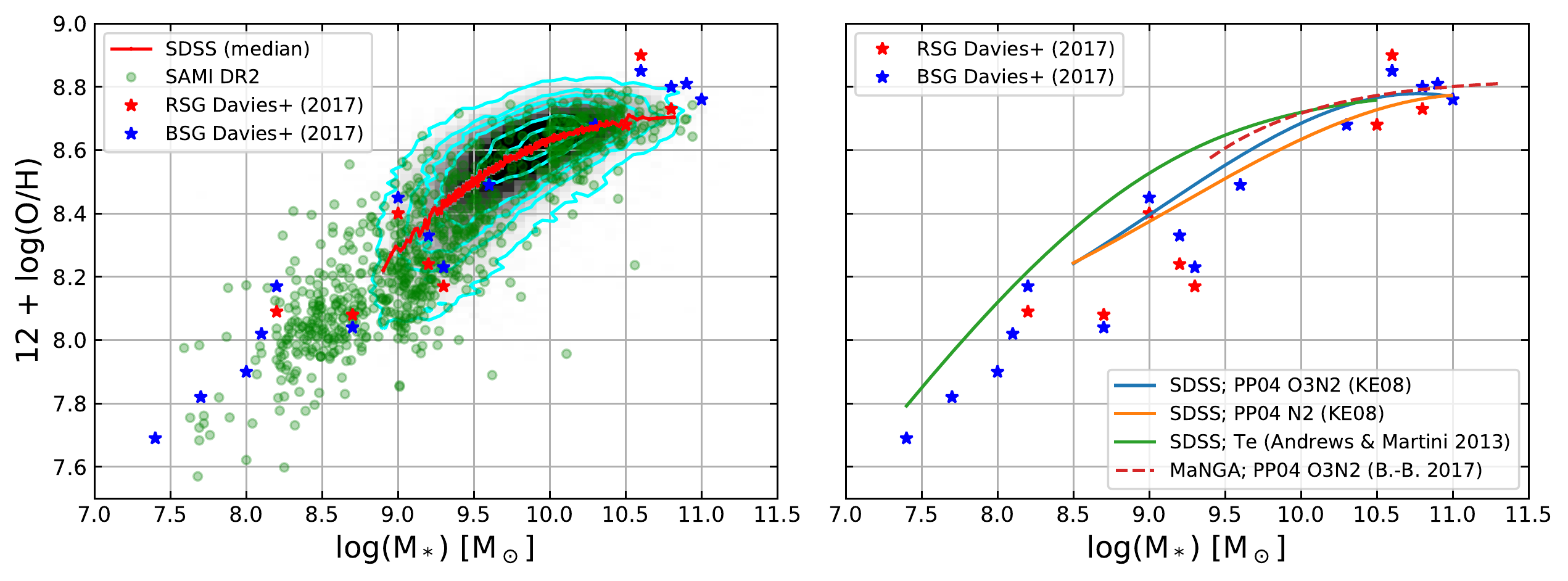}
\caption{Mass-metallicity relationships derived from various methods and calibrations. 
{\it Left:} The black 2D histogram and the cyan contours correspond to approximately 49,000 SDSS galaxies. Their metallicities are derived from the MLP model. The green circles represent 868 SAMI DR2 galaxies where the metallicities are also from the MLP model. The red curve shows the median SDSS metallicities in mass bins of 200 galaxies per bin. {\it Right:} The green curve indicates the mass-metallicity relationship derived from 
auroral lines detected in stacked SDSS spectra \citep{Andrews:2013pd}. The blue and orange curves are from applying the PP04 calibrations to SDSS galaxies \citep{Kewley:2008qy}. The red dahsed line indicates the MaNGA mass-metallicity relationship derived by \citet{Barrera-Ballesteros:2017aa}. In both panels, the blue and red supergiant metallicities from \citet{Davies:2017aa} are presented as blue and red star symbols, respectively.}\label{fig_mzr}
\end{figure*}

In Figure~\ref{fig_m51}, a similar comparison for M51 is presented. Here, the out-of-sample regions are taken from the CHAOS survey \citep{Croxall:2015aa} that do not have auroral line detections. Additionally, 6 regions from the long-slit observations of \citet{Bresolin:1999aa} are also included (the other 7 regions with low S/N on the diagnostic lines are excluded). The MLP abundances of the out-of-sample regions appear to follow the best-fit abundance gradient derived from the in-sample CHAOS~II regions, demonstrating the robustness of the metallicity predicted by the MLP calibration.

\subsection{Mass-metallicity relationship}
I now apply the MLP calibration to integrated spectra of galaxies to reproduce the mass-metallicity relationship. This is done with large numbers of spectra from the Sloan Digital Sky Survey (SDSS; \citealt{Abazajian:2009kx}) and the SAMI Galaxy Survey \citep{Croom:2012qy,Bryant:2015bh}. The SDSS line fluxes (DR7) are taken from the MPA-JHU catalog\footnote{\url{http://www.mpa-garching.mpg.de/SDSS/DR7/}}. The sample selection follows \citet{Zahid:2013kx} that includes covering fraction ($\gtrsim20\%$), redshift ($z<0.12$), and line ratio criteria (reject AGN base on \citealt{Kauffmann:2003vn} and \citealt{Kewley:2006lr}) to ensure only star-forming galaxies remain in the sample. Galaxies with S/N less than 3 in any of the diagnostic lines are also removed. The line fluxes are extinction corrected using H$\alpha$/H$\beta$ Balmer decrement and the extinction law by \citet{Cardelli:1989qy}. In total, about 49,000 galaxies are further analysed. 

The SAMI data are taken from the data release 2 \citep{Scott:2018aa}. Aperture spectra of 1 effective radius are adopted. Extinction correction also follows the \citeauthor{Cardelli:1989qy} extinction curve using the H$\alpha$/H$\beta$ Balmer decrement. Similarly, AGNs are removed using the [\ion{N}{ii}]/H$\alpha$ and [\ion{O}{iii}]/H$\beta$ line ratios \citep{Kauffmann:2003vn,Kewley:2006lr} and a S/N criterion of 3 is imposed. In total, 868 SAMI galaxies are used. 

In the left panel of Figure~\ref{fig_mzr}, I present the mass-metallicity relationship derived from SDSS and SAMI spectra using the MLP calibration. In the right panel of Figure~\ref{fig_mzr}, I show the mass-metallicity relationships from applying the PP04 O3N2 and N2 calibrations to SDSS spectra \citep{Kewley:2008qy}. I also show the relationship derived from oxygen auroral lines measured in SDSS stacked spectra \citep{Andrews:2013pd}. In both panels, the blue and red star symbols indicate the mass-metallicity relationship derived from blue and red supergiant stars (BSGs and RSGs) in nearby galaxies (\citealt{Davies:2017aa} and references therein). 
The MLP calibration yields reasonable slope and the zero point of the mass-metallicity relationship and follows closely the BSG/RSG relationship over 1 order of magnitude in oxygen abundance.

\section{Prospect and Challenge}
I have demonstrated that by fitting literature \HII\ region data with one of the simplest form of neural network, i.e.~the multi-layer perceptron, it is possible to calibrate a strong-line diagnostic that performs reasonably well over a wide range of oxygen abundance. The MLP calibration performs equally well as the R and S calibrations by \citet{Pilyugin:2016aa} and out performs some popular literature calibrations base on N2 and O3N2. 

The very first, simple attempt of this methodology demonstrates the prospect of applying this approach to large datasets coming in the near future. In particular, Figures~\ref{fig_explore_params_2} and \ref{fig_explore_params_1} suggest that with larger training samples complex models can be developed and more accurately predict oxygen abundance. Although in this work I only develop a calibration requiring five strong forbidden lines, customised calibrations can be easily developed and their performance quantified by following the same work-flow. 

The non-linear nature of the neural network, to an arbitrarily complex degree, delivers several advantages for developing strong-line diagnostics. It is no longer critical to navigate through high-dimensional space via a series of 2-dimensional projections. For example, it is common practice to first select a primary feature sensitive to metallicity and search in the residuals for potential correlations with other line ratios. Non-linear correlation  can also be more easily captured. For example, existing calibrations using the $\rm R_{23}$ index sometimes provide two different fits for the upper and lower branches due to the turn-over of the $\rm R_{23}$ index with metallicity \citep[e.g.][]{Pilyugin:2001aa,Pilyugin:2005aa}. Such upper/lower branch separation is integrated automatically in the MLP model. Furthermore, different line ratios are often sensitive to metallicity in different metallicity intervals. The MLP model adapts to this internally without needing to fit different indexes over different intervals and report the fit parameters and the applicable intervals \citep[e.g.][]{Curti:2017aa}. This makes both the development and application of a calibration more straightforward. 

In this work, no attempt was made to optimise the choice of input features. The input features are picked virtually blindly base on possible arithmetic operations of the strong lines. While some features are known to be sensitive to metallicity, others might carry only limited information. It is thus expected that the network can perform even better with an optimised input feature set. The optimisation can be guided by Figure~\ref{fig_feature_sensitivity} but will require some trial-and-error by, for example, dropping one feature at a time and re-training the network and re-tuning the hyper-parameters. This is because that the network may response to feature trimming non-linearly and many existing features are degenerate. 

Some additional features might also be important and are currently missing. For example, the electron density of the nebula, traced by either the $[\ion{S}{ii}]\lambda6716$ to $[\ion{S}{ii}]\lambda6731$ ratio or the $[\ion{O}{ii}]\lambda3726$ to $[\ion{O}{ii}]\lambda3729$ ratio, is a fundamental, yet missing physical quantity. The ionisation parameter is another important quantity that could improve metallicity calibration. Although the ionisation parameter in principle is encoded in the $[\ion{O}{iii}]\lambda5007$ to $[\ion{O}{ii}]\lambda\lambda3726,3729$ ratio already seen by the model, the ratio is also sensitive to metallicity \citep[e.g.][]{Kewley:2002fj}. The $[\ion{S}{iii}]\lambda\lambda9069,9532$ to $[\ion{S}{ii}]\lambda\lambda6716,6731$ ratio is another key feature for passing the ionisation parameter information to the model given its insensitivity to metallicity \citep[e.g.][]{Kewley:2002fj}. These line ratios are currently missing in my attempt due to their non-uniform availability in the literature. Exploring the importance of these features using coming data from, for example, the CHAOS Project may be fruitful.

Despite the prospect and advantages of the neural network over traditional calibration methods, there are limitations and challenges. The predictive power of a neural network depends heavily on the quality and quantity of the training sample. The training sample in this work has only 43 \HII\ regions (4.5\%) more metal rich than 12+log(O/H) = 8.69 (solar), only 3 regions (0.3\%) above 8.87 (1.5 solar), and no regions above 8.99 (2 solar). The lack of high metallicity training data is directly reflected in the mass-metallicity relationship (Figure~\ref{fig_mzr}). The SDSS and SAMI galaxies have no MLP oxygen abundances larger than approximately 12+log(O/H) = 8.8 or 1.3 solar. The mass-metallicity relationship from the blue and red supergiants, however, reaches up to 12+log(O/H) = 8.9. The systematic underestimate of MLP metallicities for the high-metallicity and high-mass galaxies results in the SDSS median mass-metallicity relationship to saturate at approximately 12+log(O/H) = 8.7 at the high-mass end. 
Furthermore, a low-metallicity floor of 7.4 was also imposed due to the sparse sampling in the low-metallicity end (see Section~2.1). The lack of high and low metallicitiy training data puts a fundamental constraint on the interval where the calibration is reliable. For most literature calibrations, extrapolating the calibrations outside the applicable metallicity interval can be easily visualised, and in some cases justified by comparing with photoionisation models \citep[e.g.][]{Maiolino:2008aa}. Nonetheless, the non-linear nature of the neural network puts a fundamental constraint on making predictions outside the parameter space populated by the training sample. 

\begin{figure}
\includegraphics[width = \columnwidth]{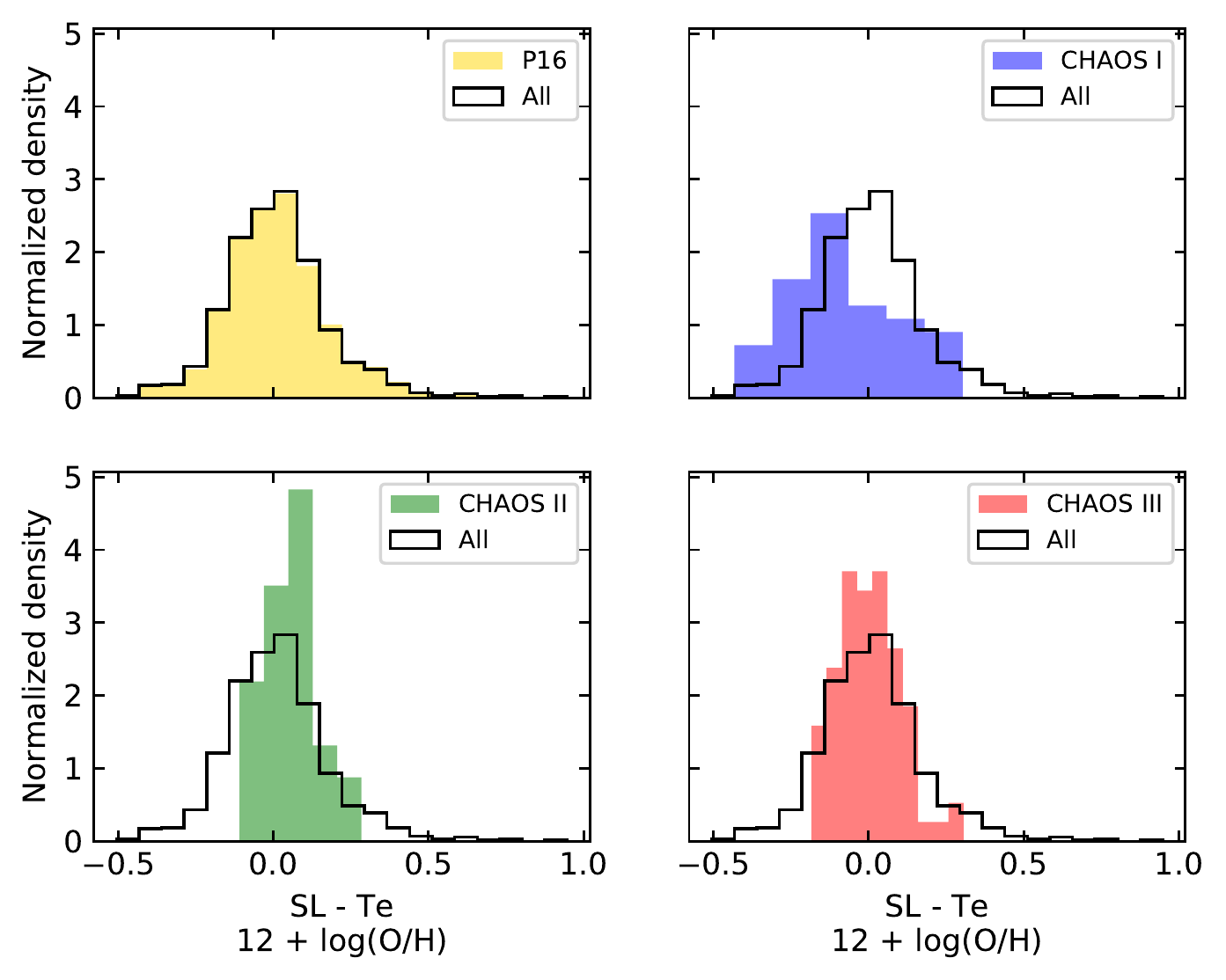}
\caption{Oxygen abundance residuals of the different samples. The residuals are defined as the abundance difference between the strong-line, MLP model and the \te\ method. The skewed distribution of CHAOS~I is discussed in Section~4.}\label{fig_compare_residuals}
\end{figure}

The uniformity of the training data is also key to the quality of the predicted metallicity. Non-uniformly analysed or low quality input data will introduce noise and bias to the network. Artefacts arose from non-uniform analysis can already been noticed in my MLP model. In Figure~\ref{fig_compare_residuals}, I present the residuals of \HII\ regions from different sources. The P16 residual distribution is very similar to the overall residual distribution, because the P16 sample comprises the majority (84\%) of the working sample. The CHAOS distributions, however, are noticeably different from the overall distribution, with CHAOS~I significantly skewed to negative residuals and CHAOS~II and III more consistent with the overall distribution. 

\begin{figure}
\includegraphics[width = \columnwidth]{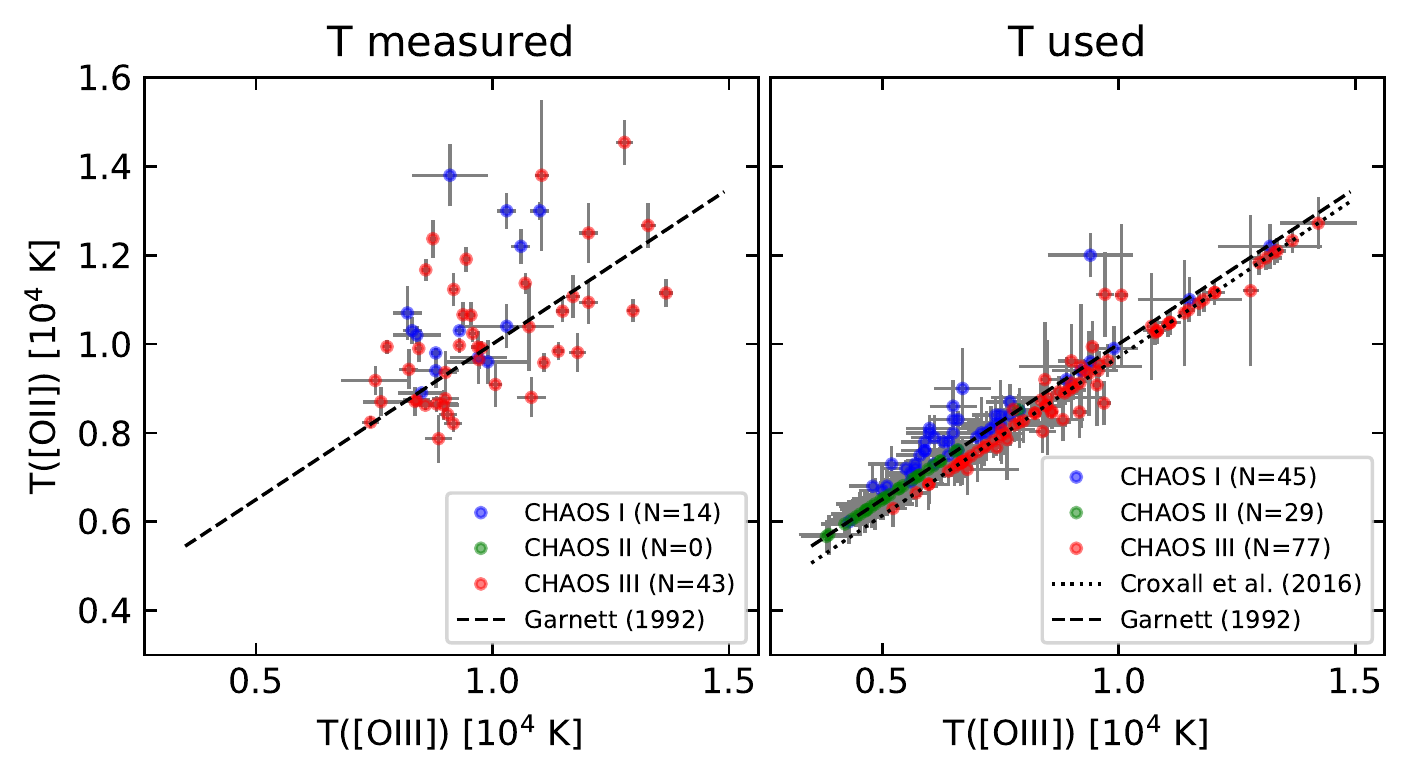}
\caption{
T([\ion{O}{ii}]) and T([\ion{O}{iii}]) of the CHAOS samples. The left panel shows the ``measured'' values and the right those ``used'' to calculate oxygen abundance. The dashed lines indicate the G92 temperature scaling relationship. The dotted line indicate the updated relationship fit by \citet{Croxall:2016tg}. }\label{fig_t2_t3_1}
\end{figure}

The differences come from how the oxygen abundances are derived from auroral line measurement(s). In the P16 sample, the temperature scaling relationship of \citet{Garnett:1992aa} is always used to infer the high- or low-state oxygen temperatures. In fact, 92\% and 19\% of the P16 sample have T([\ion{O}{iii}]) and T([\ion{N}{ii}]) measured from the [\ion{O}{iii}]$\lambda4363$ and  [\ion{N}{ii}]$\lambda5755$ lines, respectively. As in the two-zone model, T([\ion{O}{ii}]) is approximated by T([\ion{N}{ii}]). A small fraction of the sample (11\%) has both auroral lines detected and the mean of the two metallicities is adopted. In the CHAOS Project, multiple auroral lines are typically detected in each regions. Different auroral lines and scaling relationships are adopted to determine the low and high state temperatures. For the low state, T([\ion{O}{ii}]) is usually approximated by T([\ion{N}{ii}]) even when T([\ion{O}{ii}]) is measured. For the high state, CHAOS~I prioritizes T([\ion{S}{iii}]) over T([\ion{O}{iii}]) while CHAOS~III adopts the opposite. CHAOS~II has no [\ion{O}{iii}]$\lambda4363$ detections and adopts T([\ion{S}{iii}]). When only one of the two states is measured, scaling relationships are adopted. CHAOS~I and II adopt the G92 relationships, and CHAOS~III adopts updated relationships fit to their sample. These various factors determine the final T([\ion{O}{ii}]) and T([\ion{O}{iii}]) used for calculating the oxygen abundances.  Figure~\ref{fig_t2_t3_1} shows the T([\ion{O}{ii}]) and T([\ion{O}{iii}]) ``used'' (right panel) and those ``measured'' (left panel). For the CHAOS~II and III sample, the temperatures used in determining oxygen abundances lie close to the G92 scaling relationship, consistent with the P16 sample. This consistency is directly reflected in the residual distributions in Figure~\ref{fig_compare_residuals}. A significant fraction of the CHAOS~I sample, however, falls systematically above the G92 scaling relationship, resulting in the skewed residual distribution in Figure~\ref{fig_compare_residuals}.

In general, there is an unavoidable trade-off between sample uniformity and sample size. In this exploratory work, sample size is prioritized over uniformity. Thus, all the CHAOS samples are included even though a significant fraction (56\%) has no [\ion{O}{iii}]$\lambda4363$ detections. Nonetheless, it is expected that in the next few years, uniform analyses on large samples will be feasible by virtue of emerging high quality \HII\ region catalogs. The CHAOS Project has detected multiple auroral lines in a few tens \HII\ regions in each of the three published galaxies. With a sample size of a dozen nearby galaxies, close to 1,000 \HII\ regions are expected. Large IFU surveys of nearby galaxies with 10-m class telescope (e.g. the MUSE/VLT large program by the PHANGS collaboration\footnote{\url{https://sites.google.com/view/phangs/home}}; \citealt{Kreckel:2018aa}; Ho et al. in preparation) will also report auroral line metallicities. The Local Volume Mapper of the planned SDSS~V survey will detect auroral lines in thousands of \HII\ regions in M31, M33, LMC, SMC, and the Galaxy \citep{Kollmeier:2017aa}. These coming data will significantly improve the quality and quantity of the training data (even at super solar abundances), allowing both an increase of neural network complexity and its predictive accuracy.


\section*{Acknowledgements}
I thank Prof. Pilyugin for sharing the literature auroral line data. This work would not have been possible without the sample he collected through out the years. I am also deeply grateful to the support of an MPIA fellowship that enables this exploratory work. 
Additionally, I would like to thank H. Jabran Zahid, Po-Feng Wu, Rolf-Peter Kudritzki and Danielle Berg for their feedback and help at various stages of this project. 







\appendix
\section{Variation of MLP prediction due to line flux uncertainty}
\begin{figure}
\includegraphics[width = \columnwidth]{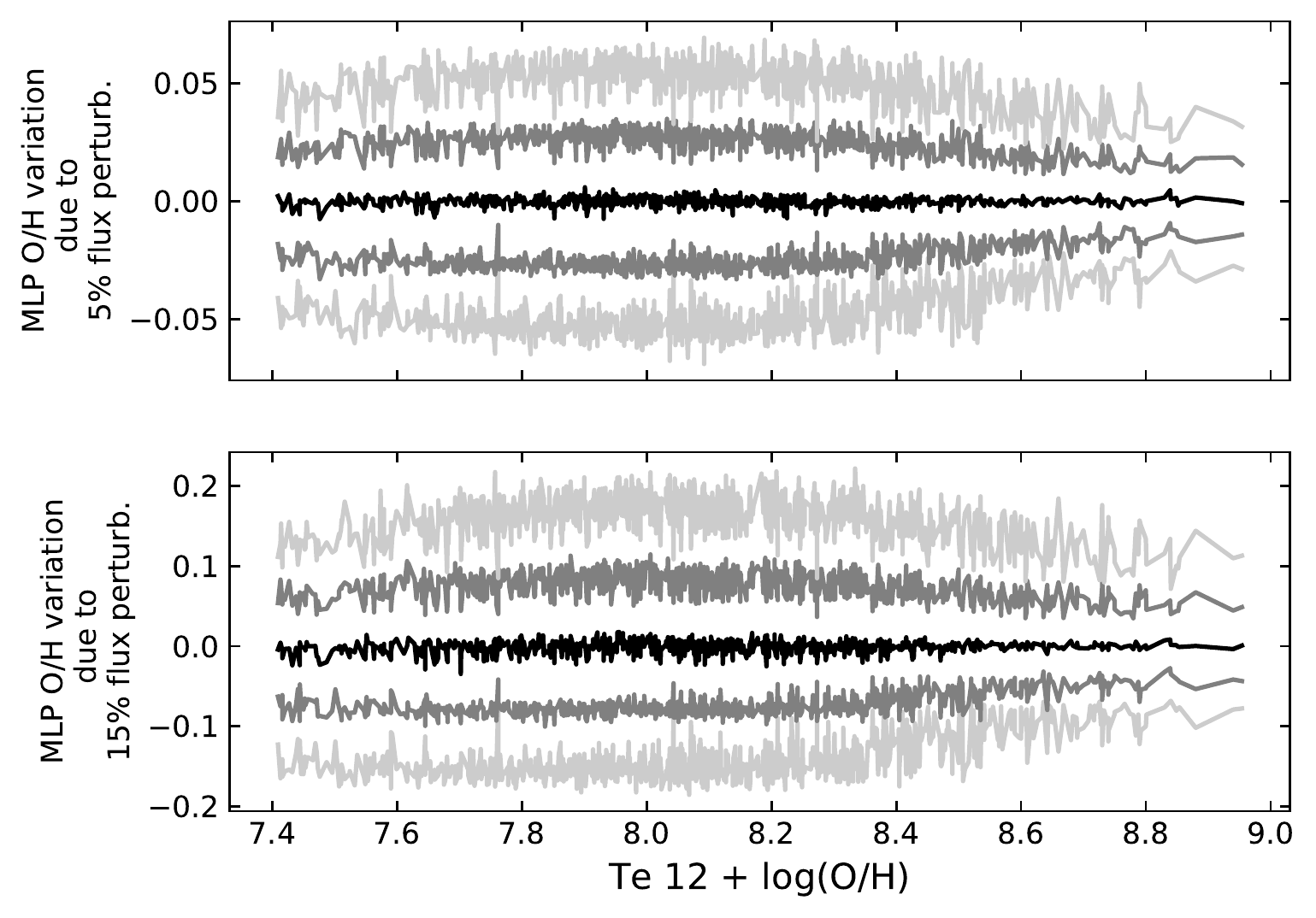}
\caption{Variation of MLP prediction due to (Gaussian) noise in the line fluxes. The top and bottom panels show two different levels of flux perturbation, 5\% (top) and 15\% (bottom). Five hundred realisations are created for each of the 962 regions. The five curves in each panel describe the distribution function of the MLP abundance fluctuation (2.3\%, 15.9\%, 50.0\%, 84.1\%, and 97.7\%). The MLP abundance fluctuation is defined as the abundance difference between predictions using noise-free and noisy line fluxes.}\label{fig_error_analysis}
\end{figure}

Non-linear models often response non-linearly with input perturbations. In the presence of measurement noise, the non-linearity could be an undesirable characteristic for calibration. To understand how the MLP model reacts to measurement uncertainties of the input line fluxes, I perform Monte Carlo simulations. Gaussian noises are added to the four line fluxes ([\ion{O}{ii}], [\ion{O}{iii}], [\ion{N}{ii}], and [\ion{S}{ii}]) and the resulting features are adopted to derive MLP abundances. Five hundred realisations are created for each of the 956 regions. Two noise levels are considered, 5\% and 15\% flux perturbations. The resulting difference between MLP abundances derived using noise-free and noisy features is shown in Figure~\ref{fig_error_analysis}. For the 5\% and 15\% perturbations, the $1\sigma$ MLP abundance fluctuations are approximately 0.025 and 0.08~dex. There is no appreciable bias in the medians. That is, adding noise to the line fluxes does not seem to skew the resulting MLP abundances in a statistically significant way.

\bibliographystyle{mnras}
\bibliography{references} 
\end{CJK*}

\bsp	
\label{lastpage}
\end{document}